\documentclass[12pt,draftclsnofoot, onecolumn]{IEEEtran}
\IEEEoverridecommandlockouts
\usepackage{amssymb}
\usepackage[cmex10]{amsmath}
\usepackage{stfloats}
\usepackage{graphicx}
\usepackage{subfigure}
\usepackage{tabularx}
\usepackage{epsfig,epsf,color,balance,cite}
\usepackage{verbatim}
\usepackage{url}
\usepackage{bm}
\usepackage{caption}
\usepackage{threeparttable}
\usepackage{diagbox}
\usepackage{multirow}

\usepackage{algorithm}
\usepackage{algorithmic}
\makeatletter

\newcommand{\Rmnum}[1]{\expandafter\@slowromancap\romannumeral #1@}
\makeatother

\definecolor{myc1}{rgb}{0,0,1}
\begin{document}

\title{Optimal Resource Allocation for Multi-UAV Assisted Visible Light Communication\\
}
\author{
	\IEEEauthorblockN{
		Yihan Cang,
		Ming Chen, \IEEEmembership{Member, IEEE},
		Zhaohui Yang, \IEEEmembership{Member, IEEE},
		Mingzhe Chen, \IEEEmembership{Member, IEEE},
		Chongwen Huang
		\vspace{-2em}
	}
	\thanks{Y. Cang and M. Chen are with the National Mobile Communications Research Laboratory, Southeast University, Nanjing 210096, China,  Emails: yihancang@seu.edu.cn, chenming@seu.edu.cn.}
	\thanks{Z. Yang  is with the Centre for Telecommunications Research, Department of Engineering, King's College London, WC2R 2LS, UK, Emails: yang.zhaohui@kcl.ac.uk.}
	\thanks{M. Chen  is  with the Department of Electrical Engineering, Princeton University, Princeton, NJ, 08544, USA, Email: mingzhec@princeton.edu.}
	\thanks{C. Huang  is  with the Zhejiang Provincial Key Lab of Information Processing, Communication and Networking, Zhejiang University, Hangzhou 310027, China, Email: chongwenhuang@zju.edu.cn.}
}

\maketitle

\begin{abstract}
In this paper, the optimization of deploying unmanned aerial vehicles (UAVs) over a reconfigurable intelligent surfaces (RISs)-assisted visible light communication (VLC) system is studied.  In the considered model, UAVs are required to simultaneously provide wireless services as well as illumination for ground users. 
To meet the traffic and illumination demands of the ground users while minimizing the energy consumption of the UAVs, one must optimize  UAV deployment, phase shift of RISs,  user association and RIS association.  This problem is formulated as an optimization problem whose goal is to minimize the transmit power of UAVs via adjusting UAV deployment, phase shift of RISs, user association and RIS association. To solve this problem, the original optimization problem is divided into four subproblems and an alternating algorithm is proposed. Specifically, phases alignment method and semidefinite program (SDP) algorithm are proposed to optimize the phase shift of RISs. Then,  the UAV deployment optimization is solved by the successive convex approximation (SCA) algorithm. Since the problems of user association and RIS association  are integer programming, the fraction relaxation method  is adopted before using dual method to find the optimal solution. For simplicity, a greedy algorithm is proposed as an alternative to optimize RIS association. The proposed two schemes demonstrate the superior performance of $34.85\%$ and $32.11\%$ energy consumption reduction over the case without RIS, respectively, through extensive numerical study. 
\end{abstract}

\begin{IEEEkeywords}
Visible light communication, unmanned aerial vehicles, reconfigurable intelligent surfaces, energy consumption.
\end{IEEEkeywords}

\section{Introduction}
Nowadays, mobile applications such as mobile health computing,
mobile object recognition and extended reality
 are emerging\cite{8245811,8014294}. Their requirements such as extremely high data rate, high quality-of-service (QoS),  and ultra-low latency are driving the revolution of mobile wireless network\cite{9170653}.  Fortunately, reconfigurable intelligent surfaces (RISs) or intelligent reflecting surfaces (IRSs) are envisioned as one of the most promising and revolutionizing technologies for improving the spectrum and energy efficiency in wireless systems\cite{8910627}.  
An RIS that consists of an array of passive reflecting elements can propagate the received signals towards the receiver by adjusting the phase shift of each reflecting element\cite{pan2020reconfigurable}. Hence, one can flexibly  enhance or weaken the signals at the receiver via adjusting the elements of RISs. Meanwhile, since the RIS reflecting elements only
passively reflect the incoming signals without any signal processing
(SP) operations, RISs can use much less power for signal transmission  compared with relays.
	
Recently, a number of existing literature such as \cite{9090356,9201413,yang2020beamforming,9027303,9247583,9217565,9217212,9133094,9217117,9234527,9148537,9224676,9110888,9148781,9238887} have focused on the applications of RISs in wireless communication. In \cite{9090356}, the authors maximized the weighted sum-rate in RIS-aided multi-cell networks.  The coverage of a downlink RIS-assisted network that consists of one base station (BS) and one user was analyzed and maximized in \cite{9201413}. The work in \cite{yang2020beamforming} optimized the resource allocation in a network that consists of a RIS-assisted wireless transmitter and multiple receivers.  The performance of RIS-assisted nonorthogonal-multiple-access (NOMA) system is analyzed in \cite{9027303,9247583,9217565,9217212,9133094}. The aforementioned works in \cite{9090356,9201413,yang2020beamforming,9027303,9247583,9217565,9217212,9133094,9217117,9234527} studied the application of RISs in radio frequency (RF) communication. Meanwhile, the works \cite{9148537} and \cite{9224676,9110888,9148781,9238887} studied the application of RISs in terahertz (THz) band and millimeter wave band, respectively. However, none of these existing works \cite{9090356,9201413,yang2020beamforming,9027303,9247583,9217565,9217212,9133094,9217117,9234527,9148537,9224676,9110888,9148781,9238887} studied the use of RISs for visible light communication (VLC) system.

VLC that utilizes the indensity of light to carry information has become an prevalent development trend in the future indoor scene due to the increasing shortage of radio spectrum resources\cite{6685758,6963803,6072221}. VLC has advantages over  RF on the aspects of huge bandwidth, excellent energy efficiency, no health hazards\cite{7239528,7072557}.
Moreover, compared with RIS-aided RF, RIS-aided VLC can provide communications and illumination simultaneously.  
However, the transmitted signals must be real and non-negative in VLC, thus the channel capacity in VLC is different from that in RF. Therefore, the aforementioned conventional methods in  RIS-aided RF \cite{9090356,9201413,yang2020beamforming,9027303,9247583,9217565,9217212,9133094,9217117,9234527,9148537,9224676,9110888,9148781,9238887} can not be directly employed in VLC due to its unique characteristics.  

Furthermore, to enable VLC to be utilized in outdoor scenario, one can use  unmanned aerial vehicles (UAVs) to provide both communications and illumination to ground users. A number of existing works has studied the problems related VLC-enabled UAVs. In particular,  in \cite{8715400}, the power consumption of VLC-enabled UAVs that must provide communications and illumination is optimized. Authors in \cite{9140367} utilized  machine learning (ML) technology to  predict the illumination requirements of users so as to optimize the deployment of the VLC-enabled UAVs. In \cite{9075277},  the authors studied the use of NOMA techniques for VLC-enabled UAVs to maximize the sum rate of all users.  
However, the Optical Wireless Channel (OWC) consists of Line-of-Sight (LOS) channel, which represents the rectilinear propagation between  transmitter and receiver, and Non-Line-of-Sight (NLOS) channel, which fades severely during the propagation via reflection, scattering and so on\cite{5682214,8301878}. In outdoor scenario, there exists a lot of obstacles such as buildings, large billboard and even trees. With the help of RIS, a UAV-RIS-user link which consists of two LOS sublinks can be constructed even there exists obstacles between UAV and ground users. 


The main contribution of this work is a novel framework that enables the UAVs to jointly use RIS and VLC to efficiently serve ground users. The key contributions are listed as follows:
\begin{itemize}
	\item The optimization of deploying UAVs over a  RISs-assisted visible light communication (VLC) system is studied. These UAVs must simultaneously provide  communications and illumination for ground users. With the constraints of data rate and illumination demand, a mixed integer programming probem is formulated which jointly optimizes UAV deployment, phase shifts of RISs,  user association and RIS association so as to minimize the  transmit power of UAVs.   
	\item To solve this problem, an algorithm that alternately optimizes UAV deployment, phase shifts of RISs,  user association and RIS association is proposed. In particular, first, phases alignment method and SDP algorithm are proposed to optimize the phase shift of RISs in two application scenarios, i.e. only one user is associated with one UAV and more than one users are associated with one
	UAV, respectively. Then, the noncave UAV deployment optimization problem is transformed to a convex problem which can be solved by CVX toolboox. Since the
	problems of user association and RIS association are integer programming, the fraction relaxation method
	is adopted before using dual method to find the optimal solution. For simplicity, a greedy algorithm
	is proposed as an alternative to optimize RIS association. 
\end{itemize}

 Through extensive numerical study, the proposed two schemes demonstrate the superior performance of $34:85\%$ and $32:11\%$ energy consumption reduction over the case without RIS, respectively. Our results also show that by associating each RIS with the closest UAV, one can achieve the minimum transmit power of all the UAVs.

	The remainder of this paper is organized as follows. The system model and problem formulation are described in Section \uppercase\expandafter{\romannumeral2}. 
	The joint UAV deployment and resource allocation is presented in Section \uppercase\expandafter{\romannumeral3}. 
	Simulation results are analyzed in Section \uppercase\expandafter{\romannumeral4}. Conclusions are drawn in Section \uppercase\expandafter{\romannumeral5}.

\section{System Model}
\begin{figure}[t]
		\centering
		\includegraphics[width=5in, height=3in]{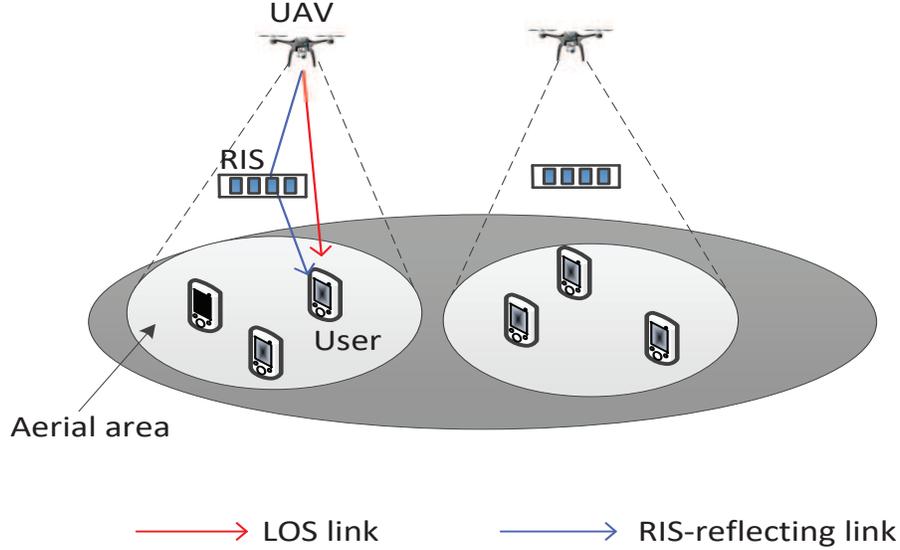}
		\caption{The architecture of a VLC-enabled UAV mutilcell network with RIS-aided.} \label{model}
	\end{figure}

	Consider a VLC-enabled UAV network which consists of a set $\mathcal{D}$ of $D$ UAVs in a specific area $\mathcal{A}$ with a set $\mathcal{L}$ of $L$ reconfigurable intelligent 
	surfaces (RISs), as shown in Fig. \ref{model}.  In this model, each UAV must simultaneously provide communication and illumination to ground users. For communication service, each UAV can directly transmit the data to the ground users or it can transmit the signal to RISs that will forward the data to the ground users. Hereinafter, we use  \emph{aerial area} to refer to the service area of each UAV. Note that each UAV does not serve ground users until it moves to the optimal location. Thus, the UAVs can be seen as static aerial base station during wireless transmission. 

	\subsection{Transmission Model}
  For simplicity, the time of each UAV flying from one place to anothor is ignored. The rotary-wing UAV is considered in this paper and each UAV can hover over one specific location to serve the users. At time slot $t$, consider a ground user $j \in \mathcal{U}$ located at $(v_{j},w_{j},0) \in \mathcal{A}$ and a flying UAV $i \in \mathcal{D}$ located at $(x_{i},y_{i},H)$, where $H$ is the  altitude of each UAV, which is assumed to be equal and fixed for all UAVs. In our model, we consider two types of data transmission: UAV-ground users and UAV-RIS-ground users. 
  The LOS channel gain between UAV $i$ and user $j$ can be expressed as:
	\begin{equation} \label{2a1}
		h_{ij}^{LOS}=\begin{cases}
			\frac{(k+1)A}{2\pi d_{ij}^2}\cos^{k}(\phi _{ij})g(\varphi_{ij})\cos(\varphi_{ij}),&0 \leq \varphi_{ij} \leq \Psi_{c},\\
			0,&\varphi_{ij} \textgreater \Psi_{c},\\
		\end{cases}
	\end{equation}
	where $k=-\frac{\ln 2}{\ln \cos(\Phi_{\frac{1}{2}})}$ denotes the Lambertian emission order with $\Phi_{\frac{1}{2}}$ being the semi angle at half-power of the transmitter.
	 $A$ denotesis the physical area
	 of the PD in each receiver and $d_{ij}$ represents the distance between UAV $i$ and ground user $j$. In \eqref{2a1}, $\phi_{ij}$ and $\varphi_{ij}$ represent the light emission angle and the incidence angle from  UAV $i$ to ground user $j$, respectively. 
	 In \eqref{2a1}, $\Psi_{c} \leq \pi /2$ denotes the field of view of the receiver, and the gain of the optical concentrator $g(\varphi_{ij})$ is defined as:
	\begin{equation}  \label{2a3}
		g(\varphi_{ij})=\begin{cases}
			\frac{n^2}{\sin^2(\Psi_{c})},&0 \leq \varphi_{ij} \leq \Psi_{c},\\
			0,&\varphi_{ij} \textgreater \Psi_{c},\\
		\end{cases}
	\end{equation} 
	where $n$ denotes internal reflective index. 
	From \eqref{2a3}, we can see that $g(\varphi_{ij})$ is a  constant when $0 \leq \varphi_{ij} \leq \Psi_{c}$.

	Let us consider a scene where there exists one or more RISs in the aerial area of UAV $i$. For tractabilty, let $m_{il}$ denote the association between UAV $i$ and RIS $l$.  If RIS $l$ is in the aerial area of UAV $i$ at time $t$, we have $m_{il}=1$; otherwise, 
	$m_{il}=0$. Since each RIS can be located in only one UAV's aerial area, then we have the following equation:
	\begin{equation}  \label{2a4}
		\sum_{i \in \mathcal{D}} m_{il}=1,  \ \ \forall l \in \mathcal{L}
	\end{equation}

\begin{figure}[t]
	\centering
	\includegraphics[width=3in]{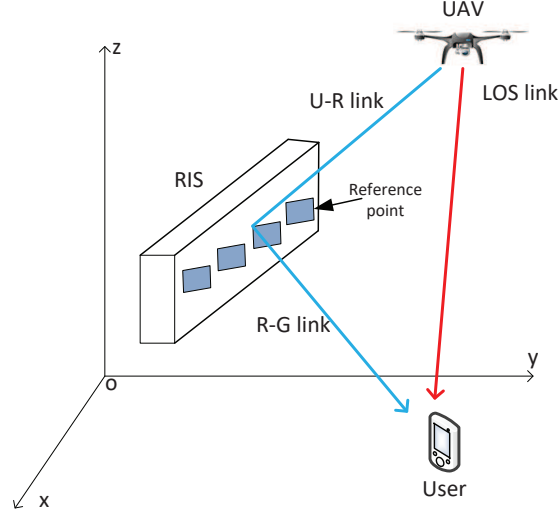}
	\caption{A RIS-assisted VLC communication system.} \label{RIS-model}
\end{figure}

	In this case, there eixts not only LOS link, but also RIS-reflecting links between UAV $i$ and ground user $j$. As shown in Fig. \ref{RIS-model}, the RISs are deployed on the buildings. The location of RIS $l$ is denoted by $(a_l,b_l,z_R)$. Without loss of generality, the height of all RISs is assumed to be the same.
	
	Assuming that all the RISs are equipped with a uniform linear array (ULA) of $M$ reflecting elements as well as a controller in the UAV to intelligently adjust the phase shifts. Denoting $\boldsymbol{\Theta}_{l}=\text{diag}\{e^{j\theta_{l1}}, e^{j\theta_{l2}}, \cdots, e^{j\theta_{lM}}\}$ as the diagonal phase-shift matrix for RIS $l$ at time $t$, where $\theta_{lm} \in [0, 2\pi), l \in \mathcal{L}, m \in \mathcal{M}=\{1, 2, \cdots, M\}$ is the phase shift of the $i$ th reflecting element of RIS $l$ at time $t$, and the phase shift $\theta_{lm}$ is assumed to be continuously controllable.

	Assuming the links from the UAV to the RIS (U-R link) and the links from the RIS to the ground user (R-G link) are both LOS channels. 
	Hence, the channel gain of the U-R link between UAV $i$ and RIS $l$ at time $t$, denoted by $\boldsymbol{h}_{il}^{UR} \in \mathbb{C}^{M \times 1}$, is given by:
	\begin{equation}  \label{2a5}
		\boldsymbol{h}_{il}^{UR}=h_{il}^{LOS}[1, e^{-j\frac{2\pi}{\lambda}d\vartheta_{il}}, \dots, e^{-j\frac{2\pi}{\lambda}(M-1)d\vartheta_{il}}]^{T}
	\end{equation}
	where the right-most term is the array reponse of an $M$-element ULA, $\vartheta_{il}=\frac{a_{l}-x_{i}}{d_{il}}$ represents the cosine of the angle of
	arrival (AoA) of the signal from UAV $i$ to the ULA at RIS $l$ at time $t$, $d$ is the antenna separation, and $\lambda$ is the carrier wavelength.
	In \eqref{2a5}, $h_{il}^{LOS}$ represents the path loss of the U-R link at time $t$ which can be expressed according to the Lambertian emission model:
	\begin{equation}  \label{2a6}
		h_{il}^{LOS}=\begin{cases}
			\frac{(k+1)A}{2\pi d_{il}^2}\cos^{k}(\phi _{il})g(\varphi_{il})\cos(\varphi_{il}),&0 \leq \varphi_{il} \leq \Psi_{c}\\
			0,&\varphi_{il} \textgreater \Psi_{c}\\
		\end{cases}
	\end{equation}
	where $d_{il}$ represents the distance between UAV $i$ and RIS $l$ at time $t$,  $\phi_{il}$ and $\varphi_{il}$ represent the light emission angle and the incidence angle from UAV $i$ to RIS $l$ at time $t$, respectively.

	Similarly, the channel gain of the R-G link between RIS $l$ and ground user $j$ at time $t$, denoted by $\boldsymbol{h}_{lj}^{RG} \in \mathbb{C}^{M \times 1}$, is given by:
	\begin{equation}  \label{2a7}
		\boldsymbol{h}_{lj}^{RG}=h_{lj}^{LOS}[1, e^{-j\frac{2\pi}{\lambda}d\vartheta_{lj}}, \dots, e^{-j\frac{2\pi}{\lambda}(M-1)d\vartheta_{lj}}]^{T}
	\end{equation}
	where  $\vartheta_{lj}=\frac{v_{j}-a_{l}}{d_{lj}}$ represents the cosine of the angle of
	departure (AoD) of the signal from RIS $l$ to ground user $j$ at time $t$, 
	$h_{lj}^{LOS}$ represents the path loss of the R-G link at time $t$ which can also be expressed according to the Lambertian emission model:
	\begin{equation}  \label{2a8}
		h_{lj}^{LOS}=\begin{cases}
			\frac{(k+1)A}{2\pi d_{lj}^2}\cos^{k}(\phi _{lj})g(\varphi_{lj})\cos(\varphi_{lj}),&0 \leq \varphi_{lj} \leq \Psi_{c}\\
			0,&\varphi_{lj} \textgreater \Psi_{c}\\
		\end{cases}
	\end{equation}
	where $d_{lj}$ represents the distance between RIS $l$ and ground user $j$ at time $t$,  $\phi_{lj}$ and $\varphi_{lj}$ represent the light emission angle and the incidence angle from RIS $l$ to ground user $j$ at time $t$, respectively.
	
	After obtaining the channel gains of both the LOS links and the RISs-reflecting links, we can further derive the  total channel gain which is the sum of channel gain of LOS link and that of all the RIS-reflecting links:
	\begin{equation}       \label{2a9}
		h_{j}(\boldsymbol{q}_{i})=|h_{ij}^{LOS}+\sum_{l=1}^{L}m_{il}(\boldsymbol{h}_{lj}^{RG})^H\boldsymbol{\Theta}_{l}\boldsymbol{h}_{il}^{UR}|
	\end{equation}
	where $\boldsymbol{q}_{i}=(x_{i},y_{i})$ represents the position of UAV $i$ at time $t$. 
	
	We consider the user association among multiple UAVs and users. Specifically, denote $u_{ij}$ as the association for UAV $i$ and ground user $j$ at time $t$. If $u_{ij} =1$, ground user $j$ is served by UAV $i$ at time $t$; otherwise, $u_{ij}=0$. Since each ground user can be served by only one UAV, we have the following equation:
	\begin{equation}          \label{2a10}
		\sum_{i \in \mathcal{D}} u_{ij}=1, \ \ \forall j \in \mathcal{U}
	\end{equation}

	Consider that when ground users are served by UAVs, they are all static, the mobility energy consumption of all the UAVs is not taken into account. Due to the limited energy of UAVs, their deployment must be optimized to minimize the transmit power while meeting the data rate and illumination requirements of ground users.

	\subsection{Problem Formulaion}
	
	In order to formulate the deployment problem of UAVs, first, the relationship between the transmit power of the UAV and the data rate required by the users must be obtained. Assuming the UAVs provide multicell channels to all the ground users, the required data rate for all the ground users at time $t$ can be formulated by: 
	
	\begin{equation}   \label{2b1}
		R_{t} \leq \frac{1}{2}\log_{2}\left (1+\frac{e}{2\pi}\left(\frac{\xi P_{i}h_{j}(\boldsymbol{q}_{i})}{n_{w}}\right)^{2}\right ), \ \   \forall i,j
	\end{equation}
	where $e$ is Euler number, $\xi$ is illumination response factor of transmitter, $n_{w}$ denotes the power of the additive white Gaussion noise (AWGN). In \eqref{2b1}, $P_{i}$ represents the transmit power of UAV $i$  at time $t$.   
	According to \eqref{2b1}, thus, we obtain the minimum transmit power of UAV $i$ that meets the data rate requirements of its associated users:
	\begin{equation}  \label{2b2}
		P_{i}\geq \frac{u_{ij}n_{w}\sqrt{\frac{2\pi}{e}(2^{2R_{t}}-1)}}{\xi h_{j}(\boldsymbol{q}_{i})}, \ \   \forall i,j
	\end{equation}

	With respect to the illumination requirements of ground users served by UAV $i$, we must have:
	\begin{equation}    \label{2b3}
		\xi P_{i}h_{j}(\boldsymbol{q}_{i}) \geq u_{ij}\eta_{j},\ \ \forall j \in \mathcal{U}
	\end{equation}
	where $\eta_{j}$ represents the illumination demand of ground user $j$ at time $t$.

	After giving the constrains of the data rate and illumination requirements of users, we can further formulate the deployment problem:
		\begin{subequations}  \label{2b4}
		\begin{align} 
			&\min_{\boldsymbol{q}_{i},\boldsymbol{u}_{i}, \boldsymbol{m}_{i}, \boldsymbol{\Theta}_{t}, P_{i}} \ \ \sum_{i \in \mathcal{D}}P_{i}, \tag{\theequation}\\  
			s.t.\ \ & \xi P_{i}h_{j}(\boldsymbol{q}_{i}) \geq u_{ij}\eta_{j},\ \ \forall i \in \mathcal{D}, \forall j \in \mathcal{U} \\
			& 	P_{i}\geq \frac{u_{ij}n_{w}\sqrt{\frac{2\pi}{e}(2^{2R_{t}}-1)}}{\xi h_{j}(\boldsymbol{q}_{i})}, \ \ \forall i \in \mathcal{D}, \forall j \in \mathcal{U} \\
			& \sum_{i \in \mathcal{D}} u_{ij}=1, \ \ \forall j \in \mathcal{U} \\
			&\sum_{i \in \mathcal{D}} m_{il}=1, \ \ \forall l \in \mathcal{L} \\
			& u_{ij}, m_{il} \in \{0,1\}, \ \ \forall i \in \mathcal{D}, \forall j \in \mathcal{U}, \forall l \in \mathcal{L} \\
			& \|\boldsymbol{q}_{i}-\boldsymbol{q}_{k}\|^{2} \geq d_{min},\ \ \forall i,k \in \mathcal{D}, i \neq k.
		\end{align}
		\end{subequations}
	where $\boldsymbol{u}_{i}=[u_{i1},u_{i2},\dots,u_{iU}]$ denotes the user association vector of UAV $i$, $\boldsymbol{m}_{i}=[m_{i1},m_{i2},\dots,m_{iL}]$ denotes the RIS association vector of UAV $i$, $\boldsymbol{\Theta}_{t} \in \mathbb{R}^{L \times M}$ denotes the phase shift matrix which can be expressed as following:
	\begin{equation}        \label{2b5}
		\boldsymbol{\Theta}_{t}=\begin{bmatrix}
			\theta_{11}&\theta_{12}&\cdots&\theta_{1M}\\
			\theta_{21}&\theta_{22}&\cdots&\theta_{2M}\\
			\vdots&\vdots&\ddots&\vdots\\
			\theta_{L1}&\theta_{L2}&\cdots&\theta_{LM}\\
		\end{bmatrix}
	\end{equation}
	and $d_{min}$ is the predefined minimum distance between any two UAVs. In \eqref{2b4}, the objective function denotes the sum transmit power of all UAVs. Constraint (\ref{2b4}a) represents the requirements of the illumination of ground users. The  data requirement for each ground user is given in (\ref{2b4}b). Constraint (\ref{2b4}c) indicates that each ground user can only be served by one UAV at each time slot. Each RIS can be located in only one UAV's aerial area as shown in (\ref{2b4}d).

	Assuming the data rate requirement and illumination reuqirement will not change in each time interval of ten minutes. Hence, in order to successfully meet the requirments of ground users, we must solve out the optimal deployment of UAVs at the beginning of each time interval.


\section{Optimization of UAV Deployment, User Association and Power Efficiency}
	According to the last section, the optimal deployment of each UAV at the beginning of each time interval can be calculated by solving the optimization problem formulated in \eqref{2b4}. Note that \eqref{2b4} is a mixed integer programming problem, an iterative algorithm is proposed to tackle with this problem. Specifically, we first optimize $\boldsymbol{q}_{i}$ and $\boldsymbol{\Theta}_{t}$ with fixed $\boldsymbol{u}_{i}$ and $\boldsymbol{m}_{i}$. 
	Afterwards, $\boldsymbol{u}_{i}$ and $\boldsymbol{m}_{i}$ can be optimized with fixed $\boldsymbol{q}_{i}$ and $\boldsymbol{\Theta}_{t}$.

	\subsection{Phase Shift Matrix Optimization and UAV Deployment }
	With fixed user association $\boldsymbol{u}_{i}$ and RIS association $\boldsymbol{m}_{i}$, the optimization problem \eqref{2b4} can be reduced to:
	\begin{subequations} \label{4a1}
	 \begin{align} 
	 	\min_{\boldsymbol{q}_{i}, \boldsymbol{\Theta}_{t}, P_{i}} &\sum_{i \in \mathcal{D}}P_{i},\tag{\theequation}\\  
	 	s.t.\ \ &  P_{i} \geq \frac{A_{ij}u_{ij}}{ h_{j}(\boldsymbol{q}_{i}) },\ \ \forall i \in \mathcal{D}, \forall j \in \mathcal{U} \\
	 	& \|\boldsymbol{q}_{i}-\boldsymbol{q}_{k}\|^{2} \geq d_{min},\ \ \forall i,k \in \mathcal{D}, i \neq k.
	 \end{align}
 	\end{subequations}
	where $A_{j}=\max\{\frac{\eta_{j}}{\xi}, \frac{n_{w}\sqrt{\frac{2\pi}{e}(2^{2R_{t}}-1)}}{\xi }\}$.
	Problem \eqref{4a1} can be solved in two steps: passive beamforming optimization and UAV deployment optimization.
	
	\subsubsection{Passive Beamforming Optimization}
	In order to futher reduce the computation complexity in passive beamforming optimization, we divide this subproblem in two application scenarios, i.e. only one user is associated with one UAV and more than one users are associated with one UAV.
	\paragraph{Only one user is associated with UAV $i$}
	In this case, it is obviously that in order to maximize the received signal energy, we can align the phases of the received signal at the ground user.  Firstly, with fixed $\boldsymbol{q}_{i}$, the total channel gain $h_{j}(\boldsymbol{q}_{i})$ in \eqref{2a9} can be further expressed as:
	\begin{align}  \label{4a2}
		h_{j}(\boldsymbol{q}_{i})=&\left |h_{ij}^{LOS}+\sum_{l \in \mathcal{L}_i}h_{lj}^{LOS}h_{il}^{LOS}\sum_{m=1}^{M}e^{j(\theta_{lm}+\frac{2\pi}{\lambda}d(m-1)(\vartheta_{lj}-\vartheta_{il}))}\right|
	\end{align}
	where $\mathcal{L}_i=\{l\in\mathcal{L}|m_{il}=1\}$ denotes the set of RISs associated with UAV $i$.
	Therefore, we can combine the signals from different paths coherently at ground user $j$, i.e., $\theta_{l1}=\theta_{l2}+\frac{2\pi}{\lambda}d(\vartheta_{lj}-\vartheta_{il})=\cdots=\theta_{lM}+\frac{2\pi}{\lambda}d(M-1)(\vartheta_{lj}-\vartheta_{il})=0,  \forall l \in\mathcal{L}_i$, or re-expressed as:
	\begin{equation}    \label{4a3}
		\theta_{lm}=\frac{2\pi(m-1)d}{\lambda}(\phi_{il}-\phi_{lj}), \ \ \forall l \in \mathcal{L}_i, \forall m \in \mathcal{M}.
	\end{equation}
	In this way, the received signal energy is maximized through the phase alignment of the 
	received signal. Hence, $h_{j}(\boldsymbol{q}_{i})$ can be further written as:
	\begin{equation}  \label{4a4}
		h_{j}(\boldsymbol{q}_{i})=\left|h_{ij}^{LOS}+\sum_{l=1}^{L}h_{lj}^{LOS}h_{il}^{LOS}M\right|
	\end{equation}
	
	\paragraph{More than one users are associated with UAV $i$}
	When an UAV is serving more than one ground users, the aligned phase $\boldsymbol{\Theta}_{t}$ will vary from one user to another if utilizing phase alignment method, thus making it a trouble to tackle with problem \eqref{4a1}. With fixed $\boldsymbol{u}_{i}$, $\boldsymbol{m}_{i}$ and $\boldsymbol{q}_{i}$ in \eqref{4a1}, we can formulate the following problem for each UAV $i\ \ (i \in \mathcal{D})$:
	\begin{equation} \label{4a5}
		\begin{aligned}
			\min_{\overset{\theta_{lm}}{l\in\mathcal{L}_i,m\in\mathcal{M}}} \max_{j\in\mathcal{U}_i}\ \ & \frac{A_{j}^{2}u_{ij}^{2}}{\left|h_{ij}^{LOS}+\sum_{l\in \mathcal{L}_i}(\boldsymbol{h}_{lj}^{RG})^H\boldsymbol{\Theta}_{l}\boldsymbol{h}_{il}^{UR}\right|^{2}}\\
			s.t. \ \ & \theta_{lm} \in [0, 2\pi), \ \ l\in\mathcal{L}_i,m\in\mathcal{M}
		\end{aligned}
	\end{equation}
	where $\mathcal{U}_i=\{j\in\mathcal U|u_{ij}=1\}$ denotes the set of ground users associated with UAV $i$.
	Problem \eqref{4a5} is a nonlinear fractional programming which can be approximated as the following problem:
	\begin{equation} \label{4a6}
		\begin{aligned}
			\min_{\substack{\theta_{lm}\\l\in\mathcal{L}_i,m\in\mathcal{M}}} \max_{j \in \mathcal{U}_i}\ \ & -\frac{\left|h_{ij}^{LOS}+\sum_{l\in\mathcal{L}_i}(\boldsymbol{h}_{lj}^{RG})^H\boldsymbol{\Theta}_{l}\boldsymbol{h}_{il}^{UR}\right|^{2}}{A_{j}^{2}u_{ij}^{2}} \\
			s.t. \ \ & \theta_{lm} \in [0, 2\pi), \ \  l\in\mathcal{L}_i,m\in\mathcal{M}
		\end{aligned}
	\end{equation}
	To solve problem \eqref{4a6}, we utilize semidefinite program (SDP) algorithm. Denote $z_{lm}=e^{-j\theta_{lm}}, \boldsymbol{z}_{l}=[z_{l1},\cdots,z_{lM}], \forall l,m$. Then $\boldsymbol{Z}_{i} \in \mathbb{C}^{|\mathcal{L}_i| \times M}$ represents the matrix of all the $\boldsymbol{z}_{l}$ that satisfy $l \in \mathcal{L}_i$ arranging in rows, where $|\mathcal{L}_i|$ represents the number of elements in set $\mathcal{L}_i$. The constraint in \eqref{4a6} is equal to $|z_{lm}|^{2}=1, l\in\mathcal{L}_i,m\in\mathcal{M}$. Through introducing $\boldsymbol{\Phi}_{ilj}=\text{diag}((\boldsymbol{h}_{lj}^{RG})^{H})\boldsymbol{h}_{il}^{UR} \in \mathbb{C}^{M}$, $(\boldsymbol{h}_{lj}^{RG})^{H}\boldsymbol{\Theta}_{l}\boldsymbol{h}_{il}^{UR}$ can be transformed to $\boldsymbol{z}_{l}^{*}\boldsymbol{\Phi}_{ilj}$. Then, the optimization problem \eqref{4a6} can be further written as:
	\begin{equation} \label{4a7}
		\begin{aligned}
			\min_{\boldsymbol{Z}_{i}} \max_{j \in \mathcal{U}_i}\ \ & -\frac{1}{A_{j}^{2}u_{ij}^{2}} \left|h_{ij}^{LOS}+\sum_{l\in\mathcal{L}_i}\boldsymbol{z}_{l}^{*}\boldsymbol{\Phi}_{ilj}\right|^{2}\\
			s.t. \ \ & |\boldsymbol{z}_{lm}|^{2}=1,\ \  l\in\mathcal{L}_i,m\in\mathcal{M}
		\end{aligned}
	\end{equation}
	Since $h_{ij}^{LOS}+\sum_{l=1}^{L}\boldsymbol{z}_{l}^{*}\boldsymbol{\Phi}_{ilj}$ is a scalar, then we can obtain:
	\begin{align}     \label{4a8}
		&\left|h_{ij}^{LOS}+\sum_{l=1}^{L}\boldsymbol{z}_{l}^{*}\boldsymbol{\Phi}_{ilj}\right|^{2}= (h_{ij}^{LOS}+\sum_{l=1}^{L}\boldsymbol{z}_{l}^{*}\boldsymbol{\Phi}_{ilj})^{H}(h_{ij}^{LOS}+\sum_{l=1}^{L}\boldsymbol{z}_{l}^{*}\boldsymbol{\Phi}_{ilj})
	\end{align}
	
	Denote $\boldsymbol{\Phi}_{ij}^{(o)(p)}=\boldsymbol{\Phi}_{oj}\boldsymbol{\Phi}_{pj}^{H}$,  $\hat{\boldsymbol{\Phi}}_{ioj}=h_{ij}^{LOS}\boldsymbol{\Phi}_{oj}$,  $\forall o,p \in \mathcal{L}_i$.
	Problem \eqref{4a7} can be further expressed as:
	\begin{subequations} \label{4a9}
		\begin{align} 
			\min_{\hat{\boldsymbol{z}}_{i}} \max_{j \in \mathcal{U}_i}\ \ & -\frac{1}{A_{j}^{2}u_{ij}^{2}} \left((h_{ij}^{LOS})^{2}+ \hat{\boldsymbol{z}}_{i}^{H}\boldsymbol{Q}_{ij}\hat{\boldsymbol{z}}_{i}\right)\tag{\theequation} \\
			s.t. \ \ & \left|[\hat{\boldsymbol{z}}_{i}]_n\right|^{2}=1, n=1,2,\cdots,|\mathcal{L}_i|M+1.
		\end{align}
	\end{subequations}

	where
		\begin{equation}       \label{4a10}
			\boldsymbol{Q}_{ij}=\begin{bmatrix}
				\boldsymbol{\Phi}_{ij}^{(1)(1)}&\boldsymbol{\Phi}_{ij}^{(1)(2)}&\cdots&\boldsymbol{\Phi}_{ij}^{(1)(|\mathcal{L}_i|)}&\hat{\boldsymbol{\Phi}}_{i1j}\\
				\boldsymbol{\Phi}_{ij}^{(2)(1)}&\boldsymbol{\Phi}_{ij}^{(2)(2)}&\cdots&\boldsymbol{\Phi}_{ij}^{(2)(|\mathcal{L}_i|)}&\hat{\boldsymbol{\Phi}}_{i2j}\\
				\vdots&\vdots&\ddots&\vdots&\vdots\\
				\boldsymbol{\Phi}_{ij}^{(|\mathcal{L}_i|)(1)}&\boldsymbol{\Phi}_{ij}^{(|\mathcal{L}_i|)(2)}&\cdots&\boldsymbol{\Phi}_{ij}^{(|\mathcal{L}_i|)(|\mathcal{L}_i|)}&\hat{\boldsymbol{\Phi}}_{i|\mathcal{L}_i|j}\\
				\hat{\boldsymbol{\Phi}}_{i1j}^{H}&\hat{\boldsymbol{\Phi}}_{i2j}^{H}&\cdots&\hat{\boldsymbol{\Phi}}_{i|\mathcal{L}_i|j}^{H}&0\\
				\end{bmatrix}
		\end{equation}
	\begin{equation}          \label{4a11}
		\hat{\boldsymbol{z}}_{i}=\begin{bmatrix}
			\boldsymbol{z}_{1}^{T}\\
			\boldsymbol{z}_{1}^{T}\\
			\vdots\\
			\boldsymbol{z}_{L}^{T}\\
			1\\
			\end{bmatrix}\in \mathbb{C}^{|\mathcal{L}_i|M+1}
		\end{equation}
	
	It can be inferred that $\hat{\boldsymbol{z}}_{i}^{H}\boldsymbol{Q}_{ij}\hat{\boldsymbol{z}}_{i}=tr(\hat{\boldsymbol{z}}_{i}^{H}\boldsymbol{Q}_{ij}\hat{\boldsymbol{z}}_{i})=tr(\boldsymbol{Q}_{ij}\hat{\boldsymbol{z}}_{i}\hat{\boldsymbol{z}}_{i}^{H})$, where $tr(\cdot)$ denotes matrix trace. In order to solve problem \eqref{4a9}, we denote $\hat{\boldsymbol{z}}_{i}\hat{\boldsymbol{z}}_{i}^{H}$ as $\hat{\boldsymbol{Z}}_{i} \in \mathbb{C}^{(|\mathcal{L}_i|M+1) \times (|\mathcal{L}_i|M+1)}$ and  $\hat{\boldsymbol{Z}}_{i}$ needs to satisfy $\hat{\boldsymbol{Z}}_{i} \succeq \boldsymbol{0}$ and $rank(\hat{\boldsymbol{Z}}_{i})=1$. Then, we relax this rank-one constraint to convert problem \eqref{4a9} to a convex SDP problem:
	
	\begin{subequations} \label{4a12}
		\begin{align}
			\min_{\hat{\boldsymbol{Z}}_{i}} \max_{j \in \mathcal{U}_i}\ \ & -\frac{1}{A_{j}^{2}u_{ij}^{2}} \left((h_{ij}^{LOS})^{2}+ tr(\boldsymbol{Q}_{ij}\hat{\boldsymbol{Z}}_{i})\right) \tag{\theequation}\\
			s.t. \ \ & [\hat{\boldsymbol{Z}}_{i}]_{nn}=1, n=1,2,\cdots,|\mathcal{L}_i|M+1\\
			& \hat{\boldsymbol{Z}}_{i} \succeq 0
		\end{align}
	\end{subequations}
	Problem \eqref{4a12} is a standerd convex problem, which can be effectively solved by using the well-known toolbox, such as CVX. Having obtained the solution of problem \eqref{4a12}, we use the Gaussian random solution to obtain a rank-one solution.

	\subsubsection{UAV deployment optimization}
	We further optimize the UAV deployment $\boldsymbol{q}_{i}$ with fixed $\boldsymbol{\Theta}_{t}$. Since $\|\boldsymbol{q}_{i}-\boldsymbol{q}_{k}\|^{2}$ in constraint (\ref{4a1}b) is a convex function with respect to $\boldsymbol{q}_{i}$ and $\boldsymbol{q}_{k}$, we can use the first-order Taylor expansion to convert it to a linear function with respect to $\boldsymbol{q}_{i}$ and $\boldsymbol{q}_{k}$:
	\begin{equation}   \label{4a13}
		\begin{aligned}
			\|\boldsymbol{q}_{i}-\boldsymbol{q}_{k}\|^{2} \geq  2(\boldsymbol{q}_{i}^{(r)}-\boldsymbol{q}_{k}^{(r)})^{T}(\boldsymbol{q}_{i}-\boldsymbol{q}_{k})-\|\boldsymbol{q}_{i}^{(r)}-\boldsymbol{q}_{k}^{(r)}\|^{2}, \ \ \forall i,k \in \mathcal{D}, i \neq k
		\end{aligned}
	\end{equation}
	where the superscript $(r)$ represents the variable at the previous iteration. Then, we can denote:
	\begin{align}  \label{4a14}
			g_{0}^{r}(\boldsymbol{q}_{i}-\boldsymbol{q}_{k}) \triangleq  2(\boldsymbol{q}_{i}^{(r)}-\boldsymbol{q}_{k}^{(r)})^{T}(\boldsymbol{q}_{i}-\boldsymbol{q}_{k})-\|\boldsymbol{q}_{i}^{(r)}-\boldsymbol{q}_{k}^{(r)}\|^{2}
		\end{align}

	Then, substituting \eqref{4a14} into \eqref{4a1}, problem \eqref{4a1} can be rewritten  as:
	\begin{subequations} 
		\begin{align}   \label{4a15}
			\min_{\boldsymbol{q}_{i}, P_{i}} \ \ &\sum_{i \in \mathcal{D}}P_{i}, \tag{\theequation}\\  
			s.t.\ \ &  P_{i} \geq \frac{A_{j}u_{ij}}{ h_{j}(\boldsymbol{q}_{i}) },\ \ \forall i \in \mathcal{D}, \forall j \in \mathcal{U} \\
			& g_{0}^{r}(\boldsymbol{q}_{i}-\boldsymbol{q}_{k}) \geq d_{min},\ \ \forall i,k \in \mathcal{D}, i \neq k.
		\end{align}
	\end{subequations}
	which is still a nonconvex problem due to the concave constraint (\ref{4a15}a). First, we introduce a group of new variables $\hat{h}_{ij}, \forall i \in \mathcal{D}, \forall j \in \mathcal{U}$. Then, problem \eqref{4a15} can be further expressed as:
	\begin{subequations} 
	 	\begin{align}   \label{4a16}
	 		\min_{\boldsymbol{q}_{i}, P_{i}, \hat{h}_{ij}} &\sum_{i \in \mathcal{D}}P_{i},  \tag{\theequation}\\  
	 		s.t.\ \ &  P_{i} \geq \frac{A_{j}u_{ij}}{\hat{h}_{ij}},\ \ \forall i \in \mathcal{D}, \forall j \in \mathcal{U} \\
	 		&\hat{h}_{ij} \leq h_{j}(\boldsymbol{q}_{i}),\ \ \forall i \in \mathcal{D}, \forall j \in \mathcal{U}\\
	 		& g_{0}^{r}(\boldsymbol{q}_{i}-\boldsymbol{q}_{k}) \geq d_{min},\ \ \forall i,k \in \mathcal{D}, i \neq k.
	 	\end{align}
	\end{subequations}
	where (\ref{4a16}a) is convex, but (\ref{4a16}b) is nonconvex. Substituting \eqref{2a9} into (\ref{4a16}b), we can obtain:
	\begin{align}   \label{4a17}
		\hat{h}_{ij} \leq \left|h_{ij}^{LOS}+\sum_{l=1}^{L}m_{il}(\boldsymbol{h}_{lj}^{RG})^H\boldsymbol{\Theta}_{l}\boldsymbol{h}_{il}^{UR}\right|
		=\left|h_{ij}^{LOS}+\sum_{l=1}^{L}\kappa_{ilj}^{(r)} h_{il}^{LOS}\right|
	\end{align}
	where $\kappa_{ilj}^{(r)} $ which is the coefficient of $h_{il}^{LOS}$ can be approximated by using the AoA of the signal at RIS at the previous iteration. In \eqref{4a17}, only $h_{ij}^{LOS}$ and $h_{il}^{LOS}$ are related with $\boldsymbol{q}_{i}$. Thus, we can further introduce a new group of variables $\hat{h}_{ij}^{LOS}, \hat{h}_{il}^{LOS}, \forall i,l,j$ into problem \eqref{4a16}:
	\begin{subequations} 
		\begin{align}   \label{4a18}
			&\min_{\boldsymbol{q}_{i}, P_{i}, \hat{h}_{ij}, \hat{h}_{ij}^{LOS}, \hat{h}_{il}^{LOS}} \ \  \sum_{i \in \mathcal{D}}P_{i},  \tag{\theequation}\\  
			s.t. \ \ &  P_{i} \geq \frac{A_{j}u_{ij}}{\hat{h}_{ij}},\ \ \forall i \in \mathcal{D}, \forall j \in \mathcal{U}, \\
			&\hat{h}_{ij} \leq \left|\hat{h}_{ij}^{LOS}+\sum_{l=1}^{L}\kappa_{ilj}^{(r)} \hat{h}_{il}^{LOS}\right|,\ \ \forall i \in \mathcal{D}, \forall j \in \mathcal{U},\\
			& g_{0}^{r}(\boldsymbol{q}_{i}-\boldsymbol{q}_{k}) \geq d_{min},\ \ \forall i,k \in \mathcal{D}, i \neq k,\\
			& \hat{h}_{ij}^{LOS} \leq \frac{(k+1)A}{2\pi d_{ij}^2}\cos^{k}(\phi _{ij}^{(r)})g(\varphi_{ij}^{(r)})\cos(\varphi_{ij}^{(r)}), \ \ \forall i \in \mathcal{D}, \forall j \in \mathcal{U},\\
			& \hat{h}_{il}^{LOS} \leq \frac{(k+1)A}{2\pi d_{il}^2}\cos^{k}(\phi _{il}^{(r)})g(\varphi_{il}^{(r)})\cos(\varphi_{il}^{(r)}), \ \  \forall i \in \mathcal{D}, \forall l \in \mathcal{L}.
		\end{align}         
	\end{subequations}
	where $\phi _{ij}^{(r)}, \varphi_{ij}^{(r)}$ represents the emission angle and the incidence angle at the previous iteration, respectively. Due to constraint (\ref{4a18}b), (\ref{4a18}d) and (\ref{4a18}e) are still nonconvex, we also use the first-order Taylor expansion method as that in \eqref{4a13}:
	\begin{subequations} 
		\begin{align}   \label{4a19}
			&\min_{\boldsymbol{q}_{i}, P_{i}, \hat{h}_{ij}, \hat{h}_{ij}^{LOS}, \hat{h}_{il}^{LOS}} \ \  \sum_{i \in \mathcal{D}}P_{i},  \tag{\theequation}\\  
			s.t. \ \ &  P_{i} \geq \frac{A_{j}u_{ij}}{\hat{h}_{ij}},\ \ \forall i \in \mathcal{D}, \forall j \in \mathcal{U}, \\
			& g_{0}^{r}(\boldsymbol{q}_{i}-\boldsymbol{q}_{k}) \geq d_{min},\ \ \forall i,k \in \mathcal{D}, i \neq k,\\
			&g_{1}^{r}(\hat{h}_{ij}^{LOS},\hat{h}_{il}^{LOS}) \geq \hat{h}_{ij}^{2},\ \ \forall i \in \mathcal{D}, \forall j \in \mathcal{U},\\
			& g_{2}^{r}(\hat{h}_{ij}^{LOS}) \geq d_{ij}^{2}, \ \ \forall i \in \mathcal{D}, \forall j \in \mathcal{U},\\
			&  g_{3}^{r}(\hat{h}_{il}^{LOS}) \geq d_{il}^{2}, \ \  \forall i \in \mathcal{D}, \forall l \in \mathcal{L}.
		\end{align}         
	\end{subequations}
	where 
		\begin{align}       \label{4a20}
			g_{1}^{r}(\hat{h}_{ij}^{LOS},\hat{h}_{il}^{LOS})=&2\text{\footnotesize Re}\left\{\left((\hat{h}_{ij}^{LOS})^{(r)}+\sum_{l=1}^{L}\kappa_{ilj}^{(r)} (\hat{h}_{il}^{LOS})^{(r)}\right)\left(\hat{h}_{ij}^{LOS}+\sum_{l=1}^{L}\kappa_{ilj}^{(r)} \hat{h}_{il}^{LOS}\right)\right\} \nonumber \\
			&-\left|(\hat{h}_{ij}^{LOS})^{(r)}+\sum_{l=1}^{L}\kappa_{ilj}^{(r)} (\hat{h}_{il}^{LOS})^{(r)}\right|^{2},\ \ \forall i \in \mathcal{D}, \forall j \in \mathcal{U}
		\end{align}
		\begin{align}       \label{4a21}
			g_{2}^{r}(\hat{h}_{ij}^{LOS})=&\frac{(k+1)A\left[2\times (\hat{h}_{ij}^{LOS})^{(r)}-\hat{h}_{ij}^{LOS}\right]}{2\pi\left((\hat{h}_{ij}^{LOS})^{(r)}\right)^{2}} \times \cos^{k}(\phi _{ij}^{(r)})g(\varphi_{ij}^{(r)})\cos(\varphi_{ij}^{(r)})          ,\ \ \forall i \in \mathcal{D}, \forall j \in \mathcal{U}
		\end{align}
		\begin{align}       \label{4a22}
			g_{3}^{r}(\hat{h}_{il}^{LOS})=&\frac{(k+1)A\left[2\times (\hat{h}_{il}^{LOS})^{(r)}-\hat{h}_{il}^{LOS}\right]}{2\pi\left((\hat{h}_{il}^{LOS})^{(r)}\right)^{2}} \times \cos^{k}(\phi _{il}^{(r)})g(\varphi_{il}^{(r)})\cos(\varphi_{il}^{(r)})          ,\ \ \forall i \in \mathcal{D}, \forall l \in \mathcal{D}
		\end{align}
	with all the $(\cdot)^{(r)}$ stands for the value of $(\cdot)$ at the previous iteration. 
	Problem \eqref{4a19} is now a convex problem which can be figured out the global optimal point by using CVX toolbox in MATLAB.

	\subsection{User and RIS Association Optimization}
	In the previous subsection,  we optimize $\boldsymbol{q}_{i}$ and $\boldsymbol{\Theta}_{t}$ with fixed $\boldsymbol{u}_{i}$ and $\boldsymbol{m}_{i}$.  In this subsection, We further optimize $\boldsymbol{u}_{i}$ and $\boldsymbol{m}_{i}$ with fixed $\boldsymbol{q}_{i}$ and $\boldsymbol{\Theta}_{t}$. Thus, the optimization problem \eqref{2b4} can be reduced to:
	\begin{subequations} \label{4b1}
		\begin{align} 
			\min_{\boldsymbol{u}_{i}, \boldsymbol{m}_{i}, P_{i}} &\sum_{i \in \mathcal{D}}P_{i},\tag{\theequation}\\  
			s.t.\ \ &  P_{i} \geq \frac{A_{j}u_{ij}}{ h_{j}(\boldsymbol{q}_{i}) },\ \ \forall i \in \mathcal{D}, \forall j \in \mathcal{U} \\
			& \sum_{i \in \mathcal{D}} u_{ij}=1, \ \ \forall j \in \mathcal{U} \\
			&\sum_{i \in \mathcal{D}} m_{il}=1,\ \  \forall l \in \mathcal{L} \\
			& u_{ij}, m_{il} \in \{0,1\}, \ \ \forall i \in \mathcal{D},\forall j \in \mathcal{U}, \forall l \in \mathcal{L}.
		\end{align}
	\end{subequations}
	where $A_{j}=\max\{\frac{\eta_{j}}{\xi}, \frac{n_{w}\sqrt{\frac{2\pi}{e}(2^{2R_{t}}-1)}}{\xi }\}$. Problem \eqref{4a1} cam be solved in two steps: user association optimization and RIS association optimization.

	\subsubsection{User Association Optimization}
	With fixed RIS Association $\boldsymbol{m}_{i}$, we adopt the fractional relaxation method to make this combinatorial problem tractable. In this case, $\boldsymbol{u}_{i}$ can take on any real value in $[0, 1]$.  Thus, problem \eqref{4b1} can be reformulated as:
	\begin{subequations} \label{4b2}
	 	\begin{align} 
	 		\min_{\boldsymbol{u}_{i}, P_{i}} &\sum_{i \in \mathcal{D}}P_{i},\tag{\theequation}\\  
	 		s.t.\ \ &  P_{i} \geq \frac{A_{j}u_{ij}}{ h_{j}(\boldsymbol{q}_{i}) },\ \ \forall i \in \mathcal{D}, \forall j \in \mathcal{U} \\
	 		& \sum_{i \in \mathcal{D}} u_{ij}=1, \ \ \forall j \in \mathcal{U} \\
	 		&u_{ij} \geq 0, \ \ \forall i \in \mathcal{D}, \forall j \in \mathcal{U}.
	 	\end{align}
	\end{subequations}
	Then, the dual problem of \eqref{4b2} can be given by:
	\begin{equation}   \label{4b3}
		\max_{\boldsymbol{\beta}} \ \  D(\boldsymbol{\beta})
	\end{equation}
	where  
	\begin{equation}       \label{4b4}
		D(\boldsymbol{\beta})=\left \{ \begin{aligned}
			\min_{\boldsymbol{u}_{i}, P_{i}} \ \ & L(\boldsymbol{u}_{i}, P_{i}, \boldsymbol{\beta}) \\
			s.t. \ \ \ \ &  \sum_{i \in \mathcal{D}} u_{ij}=1, \ \ \forall j \in \mathcal{U} \\
			& u_{ij} \geq 0, \ \ \forall i \in \mathcal{D}, \forall j \in \mathcal{U} \\
		\end{aligned}  \right. 
	\end{equation}
	with 
	\begin{align}       \label{4b5}
		L(\boldsymbol{u}_{i}, P_{i}, \boldsymbol{\beta}) 
		=\sum_{i \in \mathcal{D}}P_{i}+ \sum_{i \in \mathcal{D}} \sum_{j \in \mathcal{U}} \beta_{ij}\left(\frac{A_{j}u_{ij}}{ h_{j}(\boldsymbol{q}_{i}) }-P_{i}\right)
	\end{align}
	and $\boldsymbol{\beta}=\{\beta_{ij}\}$ is nonnegative relaxation variables with respect to (\ref{4b2}a). 
	In order to minimize the objective function of \eqref{4b3}, which is a linear combination of $u_{ij}$, we may let the smallest assoication coefficient corresponding to the $u_{ij}$ be 1 among all the UAVs with the given ground user $j$. Therefore, we can obtain the optimal $u_{ij}^{*}$:
	\begin{equation} \label{4b6}
		u_{ij}^{*}=\left \{  \begin{aligned}
			1, \ \ & \text{if} \ \  i=\arg \min _{i \in \mathcal{D}} \frac{\beta_{ij}A_{j}}{ h_{j}(\boldsymbol{q}_{i}) }\\
			0, \ \ & \text{otherwise.}
		\end{aligned} \right.
	\end{equation}
	 To achieve the optimal $P_{i}^{*}$ from \eqref{4b4}, we derive the first derivative with respect to $P_{i}$ considering that \eqref{4b4} is a linear problem with respect to $P_{i}$:
	 \begin{equation}       \label{4b7}
	 	\frac{\partial L(\boldsymbol{u}_{i}, P_{i}, \boldsymbol{\beta})}{ \partial P_{i}}=1-\sum_{j \in \mathcal{U}}\beta_{ij}
	 \end{equation}
	Note that the optimal $P_{i}^{*}=+\infty$ if $1-\sum_{j \in \mathcal{U}}\beta_{ij} \textless 0$. To avoid this, we must have $\sum_{j \in \mathcal{U}}\beta_{ij} \leq 1$. If there are multiple minimum indexes in satisfying $\arg \min _{i \in \mathcal{D}} \frac{\beta_{ij}A_{j}}{ h_{j}(\boldsymbol{q}_{i}) }$, we can choose any one of them. 
	
	The values of $\beta_{ij}$ can be determined by the gradient method. The updating procedure is given by:
	\begin{equation}     \label{4b8}
		\beta_{ij}=\left[\beta_{ij}^{(r)}+\rho \left(\frac{A_{j}u_{ij}^{(r)}}{ h_{j}(\boldsymbol{q}_{i}) }-P_{i}^{(r)}\right)\right]^{+}
	\end{equation}
	where $[a]^{+}=\max (a,0)$, and $\rho$ is a dynamically chosen positive step-size sequence. Thus, we can obtain the optimal $\boldsymbol{u}_{i}^{*}$ and $P_{i}^{*}$ with fixed $\boldsymbol{m}_{i}$ through optimizing primal variables and dual variables iteratively.

	From \eqref{4b6}, we can see that even though the feasible region of $u_{ij}$ is relaxed to be continuous, the optimal solution to problem \eqref{4b2} always satisfies the discrete constraints $u_{ij} \in \{0,1\}, \forall i \in \mathcal{D}, \forall j \in \mathcal{U}$. Hence, the relaxation of $u_{ij}$ does not lose optimality to the primal problem.

	\subsubsection{RIS Association Optimization}
	We further optimize $\boldsymbol{m}_{i}$ with fixed $\boldsymbol{u}_{i}$. In this subsubsection, we propose two efficient algorithms to tackle this problem.
	\paragraph{Dual Method}
	First, we formulate the following optimization problem respect to $\boldsymbol{m}_{i}$:
	\begin{subequations} \label{4b9}
		\begin{align} 
			\min_{\boldsymbol{m}_{i},\tilde{h}_{ij}, P_{i}} &\sum_{i \in \mathcal{D}}P_{i},\tag{\theequation}\\  
			s.t.\ \ &  P_{i} \geq \frac{A_{j}u_{ij}}{ \tilde{h}_{ij} },\ \ \forall i \in \mathcal{D}, \forall j \in \mathcal{U} \\
			&\tilde{h}_{ij}  \leq  \left|h_{ij}^{LOS}+\sum_{l=1}^{L}m_{il}(\boldsymbol{h}_{lj}^{RG})^H\boldsymbol{\Theta}_{l}\boldsymbol{h}_{il}^{UR}\right|,  \ \  \forall i \in \mathcal{D}, \forall j \in \mathcal{U}, \\
			&\sum_{i \in \mathcal{D}} m_{il}=1,\ \  \forall l \in \mathcal{L} \\
			&m_{il} \in \{0,1\},\ \ \forall i \in \mathcal{D}, \forall l \in \mathcal{L}.
		\end{align}
	\end{subequations}       
	where $\tilde{h}_{ij}, \forall i,j$ is a group of new introduced variables. For the optimal solution of problem \eqref{4b9}, constraint (\ref{4b9}b) will always hold with equality. Note that (\ref{4b9}b) is nonconvex, we can rewrite the right hand side of constraint (\ref{4b9}b) according to \eqref{4a8} due to the fact that $m_{il} \in \{0,1\}$:
		\begin{align}             \label{4b10}
			&\left|h_{ij}^{LOS}+\sum_{l=1}^{L}m_{il}(\boldsymbol{h}_{lj}^{RG})^H\boldsymbol{\Theta}_{l}\boldsymbol{h}_{il}^{UR}\right|^{2}  \nonumber \\
			=&C_{ij0}+\sum_{l=1}^{L}C_{ijl}m_{il}+\sum_{l=2}^{L} \sum_{v=1}^{l-1} C_{ijlv}m_{il}m_{iv}
		\end{align}
	where 
	\begin{align}     \label{4b11}
		&C_{ij0}=(h_{ij}^{LOS})^{2} \nonumber\\
		&C_{ijl}=2h_{ij}^{LOS}\text{Re}\{(\boldsymbol{h}_{lj}^{RG})^H\boldsymbol{\Theta}_{l}\boldsymbol{h}_{il}^{UR}\}+\left|(\boldsymbol{h}_{lj}^{RG})^H\boldsymbol{\Theta}_{l}\boldsymbol{h}_{il}^{UR}\right|^{2} \nonumber\\
		&C_{ijlv}=2\text{Re}\{(\boldsymbol{h}_{il}^{UR})^{H}\boldsymbol{\Theta}_{l}^{H}\boldsymbol{h}_{lj}^{RG}(\boldsymbol{h}_{vj}^{RG})^H\boldsymbol{\Theta}_{v}\boldsymbol{h}_{iv}^{UR}\}
	\end{align}
	In \eqref{4b10}, we denote $m_{il}m_{iv}$ as $E_{ilv}$. Thus, $E_{ilv}$ must satisfies:
	\begin{align}      \label{4b12}
		&E_{ilv} \geq m_{il}+m_{iv}-1, 0 \leq E_{ilv} \leq 1, \nonumber \\
		&E_{ilv} \leq m_{il}, m_{iv}, \ \ \forall i \in \mathcal{D}, \forall i,l \in \mathcal{L}, i \neq l.
	\end{align}
	Now, problem \eqref{4b9} can be reformulated as:
	\begin{subequations} \label{4b13}
		\begin{align} 
			&\min_{\boldsymbol{m}_{i},\tilde{h}_{ij}, E_{ilv} P_{i}} \ \ \sum_{i \in \mathcal{D}}P_{i},\tag{\theequation}\\  
			s.t.\ \ &  P_{i} \geq \frac{A_{j}u_{ij}}{ \tilde{h}_{ij} },\ \ \forall i \in \mathcal{D}, \forall j \in \mathcal{U} \\
			&\tilde{h}_{ij}^2  \leq  C_{ij0}+\sum_{l=1}^{L}C_{ijl}m_{il}+\sum_{l=2}^{L} \sum_{v=1}^{l-1} C_{ijlv}E_{ilv},   \ \ \forall i \in \mathcal{D}, \forall j \in \mathcal{U}, \\
			&E_{ilv} \geq m_{il}+m_{iv}-1, 0 \leq E_{ilv} \leq 1,  \ \  \forall i \in \mathcal{D}, \forall i,l \in \mathcal{L}, i \neq l, \\
			&E_{ilv} \leq m_{il}, m_{iv},  \ \ \forall i \in \mathcal{D}, \forall i,l \in \mathcal{L}, i \neq l,\\
			&\sum_{i \in \mathcal{D}} m_{il}=1,\ \  \forall l \in \mathcal{L} \\
			&m_{il} \geq 0,\ \ \forall i \in \mathcal{D}, \forall l \in \mathcal{L}.
		\end{align}
	\end{subequations}    
	where we have relaxed the integer constraints (\ref{4b9}d) with $m_{il} \in [0,1]$. Thus, problem \eqref{4b13} becomes a convex problem which can be tracked through the dual method.

	The dual problem of \eqref{4b13} is given by:
	\begin{equation}   \label{4b14}
		\max_{\boldsymbol{\tau},\boldsymbol{\gamma},\boldsymbol{\Gamma}} \ \  \hat D(\boldsymbol{\tau},\boldsymbol{\gamma},\boldsymbol{\Gamma})
	\end{equation}
	where  
	\begin{align}       \label{4b15}
		&\hat D(\boldsymbol{\tau},\boldsymbol{\gamma},\boldsymbol{\Gamma})=\\
		&\left \{ \begin{aligned}
			&\min_{\boldsymbol{m}_{i},\tilde{h}_{ij}, E_{ilv} P_{i}}  \hat L(\boldsymbol{m}_{i},\tilde{h}_{ij}, E_{ilv}, P_{i}, \boldsymbol{\tau},\boldsymbol{\gamma}, \boldsymbol{\Gamma}) \\
			s.t. \ \ &\sum_{i \in \mathcal{D}} m_{il}=1, \ \ \forall l \in \mathcal{L}, \\
			& m_{il} \geq 0, \ \ \forall i \in \mathcal{D}, \forall l \in \mathcal{L}, \\
			&0 \leq E_{ilv} \leq 1, \ \  \forall i \in \mathcal{D}, \forall i,l \in \mathcal{L}, i \neq l, \\
		\end{aligned}  \right.  \nonumber
	\end{align}
	with 
	\begin{align}       \label{4b16}
		&\hat L(\boldsymbol{m}_{i},\tilde{h}_{ij}, E_{ilv}, P_{i},\boldsymbol{\tau}, \boldsymbol{\gamma}, \boldsymbol{\Gamma})=\sum_{i \in \mathcal{D}}P_{i} -\sum_{i \in \mathcal{D}} \sum_{j \in \mathcal{U}}[\tau_{ij}(P_{i} - \frac{A_{j}u_{ij}}{ \tilde{h}_{ij} })]   \\
		&-\sum_{i \in \mathcal{D}} \sum_{j \in \mathcal{U}} [\gamma_{ij}( C_{ij0}+\sum_{l=1}^{L}C_{ijl}m_{il} +\sum_{l=2}^{L} \sum_{v=1}^{l-1} C_{ijlv}E_{ilv}-\tilde{h}_{ij} )] \nonumber \\
		&-\sum_{i \in \mathcal{D}}\sum_{l=2}^{L} \sum_{v=1}^{l-1}\left[\Gamma_{1ilv}(E_{ilv} - m_{il}-m_{iv}+1) +\Gamma_{2ilv}(m_{il}-E_{ilv})+\Gamma_{3ilv}(m_{iv}-E_{ilv})\right] \nonumber
	\end{align}
	and $\boldsymbol{\tau}=\{\tau_{ij}\}$, $\boldsymbol{\gamma}=\{\gamma_{ij}\}$, $\boldsymbol{\Gamma}=\{\Gamma_{1ilv},\Gamma_{2ilv},\Gamma_{3ilv}\}$ are nonnegative Lagrange multipliers with respect to the corresponding constraints in primal problem \eqref{4b13}.
	
	To minimize the objective function in \eqref{4b15}, which is a linear combination of $E_{ilv}$, we must let the positive coefficients corresponding to the $E_{ilv}$ be 0:
	\begin{equation} \label{4b17}
		E_{ilv}^{*}=\left \{ \begin{aligned}
			&1, \ \ \text{if} \ \  \Gamma_{1ilv} - \Gamma_{2ilv} - \Gamma_{3ilv} +\sum_{j \in \mathcal{U}} \gamma_{ij} C_{ijlv}\textgreater 0 \\
			&0, \ \ \text{otherwise}
		\end{aligned} \right.
	\end{equation}
	Due to the constraint $\sum_{i \in \mathcal{D}} m_{il}=1,  \forall l \in \mathcal{L}$, we must let the smallest association coefficient corresponding to $m_{il}$ be 1 among all the UAVs with given RIS $l$:
	 \begin{equation} \label{4b18}
	 	m_{il}^{*}=\left \{ \begin{aligned}
	 		&1, \ \ \text{if} \ \ i =  \arg \min_{i \in \mathcal{D}} C_{il} \\
	 		&0, \ \ \text{otherwise}
	 	\end{aligned} \right.
	 \end{equation}
	where 
	\begin{equation}        \label{4b19}
		C_{il}=\left\{ \begin{aligned}
			&-\sum_{j \in \mathcal{U}} \gamma_{ij}C_{ijl}- \sum_{v=2}^{L}(\Gamma_{3ilv}-\Gamma_{1ilv}),\ \ \text{if} \ \ l=1\\
			&-\sum_{j \in \mathcal{U}} \gamma_{ij}C_{ijl}- \sum_{v=1}^{l-1}(\Gamma_{2ilv}-\Gamma_{1ilv})-\sum_{v=l+1}^{L}(\Gamma_{3ilv}-\Gamma_{1ilv}),\ \  \text{if} \ \  2 \leq l \leq L-1         \\
			&-\sum_{j \in \mathcal{U}} \gamma_{ij}C_{ijl}- \sum_{v=1}^{L-1}(\Gamma_{2ilv}-\Gamma_{1ilv}),\ \ \text{if} \ \   l=L\\
		\end{aligned} \right.
	\end{equation}
	
	The optimal $\tilde{h}_{ij}$ can be obtained through setting the first derivative of \eqref{4b16} to 0:
	\begin{equation}        \label{4b20}
		\tilde{h}_{ij}=\sqrt[3]{\frac{\tau_{ij}A_{j}u_{ij}}{\gamma_{ij}}}
	\end{equation}
	Therefore, the optimal $P_{i}^{*}, \forall i \in \mathcal{D}$  takes the minimum value that satisfies  (\ref{4b13}a).

	The values of $\boldsymbol{\tau}=\{\tau_{ij}\}$,$\ \ \boldsymbol{\gamma}=\{\gamma_{ij}\}$ and  $\boldsymbol{\Gamma}=\{\Gamma_{1ilv},\Gamma_{2ilv},\Gamma_{3ilv}\}$ can be given by the sub-gradient method:
	\begin{align}            \label{4b21}
		&\tau_{ij}=[\tau_{ij}-\rho(P_{i} - \frac{A_{j}u_{ij}}{ \tilde{h}_{ij} })]^{+}\\
		&\gamma_{ij}=[\gamma_{ij}-\rho (C_{ij0}+\sum_{l=1}^{L}C_{ijl}m_{il}+\sum_{l=2}^{L} \sum_{v=1}^{l-1} C_{ijlv}E_{ilv}-\tilde{h}_{ij}^2)]^{+} \nonumber \\
		&\Gamma_{1ilv}=[\Gamma_{1ilv}-\rho(E_{ilv} - m_{il}-m_{iv}+1)]^{+}\nonumber\\
		&\Gamma_{2ilv}=[\Gamma_{2ilv}-\rho(m_{il}-E_{ilv})]^{+}\nonumber\\
		&\Gamma_{3ilv}=[\Gamma_{3ilv}-\rho(m_{iv}-E_{ilv})]^{+} \nonumber
	\end{align}
	
	\paragraph{Greedy Algorithm}
	To further decrease the complexity of  RIS association optimization, we propose a greedy algorithm  which optimizes one RIS association at one time. In particular, there is no RIS associated with any UAV at the beginning. Then, one RIS is added at each time and the RIS is associated with the UAV among all UAVs, which can minimize the total transmitting power. The details of our greedy algorithm is listed as Algorithm  \ref{algorithm 1}.
	
		\begin{algorithm}[t]
		\algsetup{linenosize=\small}
		\small
		\caption{Greedy Algorithm}
		\begin{algorithmic}[1]             \label{algorithm 1}
			\STATE {\bf Input:}  user association $\boldsymbol{u}_{i}$, phase shift matrix $\boldsymbol{\Theta}_{t}$,UAV deployment $\boldsymbol{q}_{i}$
			\STATE  {\bf Initialize:} $m_{il}=0$,$\forall i \in \mathcal{D}, \forall l \in \mathcal{L}$
			\STATE  {\bf For} $l=1:1:L$ {\bf do}
			\STATE   \hspace*{0.2in} $i^*=\arg \min_{i\in\mathcal{D}} \sum_{i\in\mathcal{D}}P_{i}$
			\STATE  \hspace*{0.2in} $m_{i^*l}=1$
			
			\STATE {\bf End} 
			\STATE {\bf Output:} RIS association $\boldsymbol{m}_{i}$.  
		\end{algorithmic}
	\end{algorithm}

	\begin{algorithm}[t]    
		\algsetup{linenosize=\small}
		\small
		\caption{Iterative Beamforming, Deployment, and Association Algorithm}
		\begin{algorithmic}[1]         \label{algorithm 2}
			\STATE {\bf Input:} ground users' locations, altitude of UAV $H$, data rate requirement $R_{t}$, illumination
			requirement $\eta_{j}$
			\STATE  {\bf Initialize:} $\boldsymbol{u}_{i}$,$\ \ \boldsymbol{m}_{i}$,$\ \ \boldsymbol{q}_{i}$
			\STATE {\bf repeat}
			\STATE  \hspace*{0.2in} Passive Beamforming  Optimization $\boldsymbol{\Theta}_{t}$:\\
			\hspace*{0.4in} {\bf if} Only one user is associated with UAV $i$:\\
			\hspace*{0.6in} {\bf then} solve \eqref{4a3};\\
			\hspace*{0.4in} {\bf else}: \\
			\hspace*{0.6in} {\bf then} solve \eqref{4a12} using \eqref{4a11}.
			\STATE  \hspace*{0.2in} UAV deployment optimization $\boldsymbol{q}_{i}$:\\
			\hspace*{0.4in}  Solve \eqref{4a19} using \eqref{4a14},\eqref{4a20}-\eqref{4a22}.
			\STATE  \hspace*{0.2in} User Association Optimization $\boldsymbol{u}_{i}$:\\
				\hspace*{0.4in}  Solve \eqref{4b6} \\
				\hspace*{0.4in}  Update $\boldsymbol{\beta}$ using \eqref{4b8}.
			\STATE  \hspace*{0.2in} RIS Association Optimization $\boldsymbol{m}_{i}$:\\
			\hspace*{0.4in} {\bf if} using Algorithm 1:\\
			\hspace*{0.6in} {\bf then} solve \eqref{4a3};\\
			\hspace*{0.9in} Solve \eqref{4b18} using \eqref{4b11},\eqref{4b19}\\
			\hspace*{0.9in} Update $\boldsymbol{\tau}$,$\boldsymbol{\gamma}$,$\boldsymbol{\Gamma}$ using \eqref{4b17},\eqref{4b20},\eqref{4b21}.\\
			\hspace*{0.4in} {\bf else}: \\
			\hspace*{0.6in} {\bf then} use Algorithm 2.\\
			\STATE {\bf until} the objective value \eqref{2b4} converges.
			\STATE {\bf Output:} $\boldsymbol{u}_{i}$,$\ \ \boldsymbol{m}_{i}$,$\ \ \boldsymbol{q}_{i}$, $\ \ \boldsymbol{\Theta}_{t}$	
		\end{algorithmic}
	\end{algorithm}

	\subsection{Complexity of the Proposed Algorithm}
	The proposed algorithms used to solve problem \eqref{2b4} is summarized in Algorithm \ref{algorithm 2}. The main complexity of the proposed Algorithm 2 lies in solving four subproblems: $\boldsymbol{\Theta}_{t}$, $ \boldsymbol{q}_{i}$, $\boldsymbol{u}_{i}$, $ \boldsymbol{m}_{i}$. For the passive beamforming optimization $\boldsymbol{\Theta}_{t}$, we consider two situations: if only one uer is associated with UAV $i$, the complexity of calculating $\theta_{lm}$ is $\mathcal{O}(LM)$ according to \eqref{4a3}; otherwise, the complexity of solving an SDP optimization problem \eqref{4a12} is $\mathcal{O}\left((LM+1)^3\right)$. For the UAV deployment optimization $\boldsymbol{q}_{i}$, we use SCA method to tackle this sub-problem, the complexity at each iteration is $\mathcal{O}(S_1^2S_2)$,  where $S_1=2DU+DL+3D$ denotes the total number of variables and $S_2=\frac{D(D-1)}{2}+3DU+DL$ represents the total number of constraints. To solve the UAV deployment problem, the number of iterations required for SCA is  $\mathcal{O}\left(\sqrt{S_1}\log_{2}(1/ \epsilon_1)\right)$, where $\epsilon_1$ is the accuracy of SCA.  For the user association optimization $\boldsymbol{u}_{i}$, the complexity of calculating \eqref{4b6} is $\mathcal{O}(DU)$. The iteration number of the user association problem can be estimated by $\mathcal{O}(1/ \sqrt{\epsilon_2})$, where $\epsilon_2$ denotes the accuracy of the dual method in this sub-problem. For the RIS association optimization $\boldsymbol{m}_{i}$, in the case of using the dual method,  the complexity of calculating \eqref{4b18} is $\mathcal{O}(DL)$. Besides, the iteration number until \eqref{4b13} converges can be expressed as $\mathcal{O}(1/ \sqrt{\epsilon_3})$, where $\epsilon_3$ represents the accuracy of the dual method adopted in this sub-problem. In the case of using greedy algorithm, the complexity is $\mathcal{O}(DL)$. The total complexities of the proposed Algorithm 2 with dual method and greedy algorithm in RIS association optimization are $\mathcal{O}\left(\hat{L} [(LM+1)^3+S_1^{2.5}S_2\log_{2}(1/ \epsilon_1)+DU/\sqrt{\epsilon_2}+DL/\sqrt{\epsilon_3}]\right)$ and $\mathcal{O}\left(\hat{L} [(LM+1)^3+S_1^{2.5}S_2\log_{2}(1/ \epsilon_1)+DU/\sqrt{\epsilon_2}+DL]\right)$, respectively, where $\hat{L}$ is the iteration number of the proposed iterative algorithm. In conclusion, the calculation of the proposed optimization algorithms only requires polynomial computational complexity. For the RIS association optimization problem, the computation complexity of greedy algorithm is smaller compared to the dual method.
	
	\section{simulation results}
	
	In this section, we perform simulations to corroborate the performance of our proposed algorithms. An 100 $\text{m}$ $\times$ 100 $\text{m}$ square area is considered with $U=6$ randomly distributed ground users.  $L=3$ RISs which has $M=5$ reflecting elements and $D=3$ UAVs are also deployed in this specific area. The required data rate for all the ground users is $R_t=25\text{bps/Hz}$ and the illumination requirements of all the ground users are generated randomly and uniformly over $[10^{-5},9 \times 10^{-5}]$. Moreover, the antenna separation $d$ equals to a half of the carrier wavelength $\lambda$. Other parameters are listed in Table \ref{table1}. Furthermore, the initial parameters are set with $\theta_{lm}=0, \forall l , m$. And the association of ground users and RISs are all generated randomly subject to (\ref{2b4}c),(\ref{2b4}d) and (\ref{2b4}e) at the beginning. 
	
	\begin{table}[htbp]
		\caption{Simulation Parameters} 
		\label{table1} 
		\centering 
		\begin{tabular}{|m{8cm}<{\centering}|m{1.5cm}<{\centering}|m{1.5cm}<{\centering}|}
			\hline
			\hline
			Parameters&Symbols&Values\\
			\hline
			Semi-angle & $\Phi_{\frac{1}{2}}$ &$80^{\circ}$\\
			\hline
			Detecting area of each receiver&$A$&$1\text{cm}^{2}$\\
			\hline
			Altitude of UAVs& $H$&$20\text{m}$\\
			\hline
			Field of view&$\Psi_{c}$&$90^{\circ}$\\
			\hline
			Internal reflective index& $n$&$4.5$\\
			\hline
			Altitude of RISs&$z_R$&$5\text{m}$\\
			\hline
			Minimum distance between two UAVs&$d_{min}$&$10\text{m}$\\
			\hline
			Illumination response factor of transmitter&$\xi$&$0.9\text{Amp./W}$\\
			\hline
			Power of  AWGN&$n_w$&$1 \times 10^{-12}$\\
			\hline
			\hline
		\end{tabular}
	\end{table}

\begin{figure}[t]
			\centering
			\includegraphics[width=4.2in]{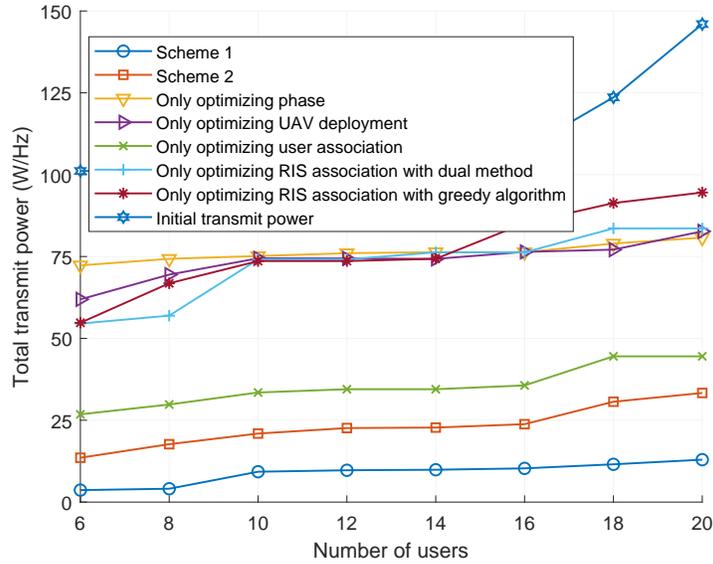}
			\caption{Total transmit power versus number of users $U$.} \label{user}
		\end{figure}

	Fig. \ref{user} illustrates the total transmit power of all the UAVs versus the number of ground users $U$ whose data requirement and illumination requirement are satified by UAVs. Scheme $\text{\Rmnum{1}}$  and Scheme $\text{\Rmnum{2}}$  represents the proposed iterative algorithm with RISs' association optimized by dual method and greedy algorithm, respectively. And the initial scheme represents the scheme with initial settings. As can be seen, our proposed two schemes both work well compared with  the initial scheme. Particularly,  Scheme $\text{\Rmnum{1}}$ outperforms Scheme $\text{\Rmnum{2}}$  due to the fact that greedy algorithm searches one optimal solution at each step while the dual method can find the global optimal solution. Meanwhile, the total transmit power increases with the increase of the number of ground users. This is because the UAVs have to satisfy the demand of all the ground users. 	In addition, the schemes that only optimize one of these four optimization variables are also shown in Fig. \ref{user}. It is obvious that these four schemes all have the ability to reduce the total transmit power of UAVs. And the scheme that only optimizes the user association outperforms the other three schemes. This indicates  that  an appropriate user association can greatly reduce the total transmit power. And it is reasonable to only optimize user association when an urgent deployment is required. Moreover, we can find that there is a flateau when the number of users growths from 10 to 16. This may be explained by that the  power needs to be provided for the increased users is less than the  power needed by the user who requires the most transmit power in the same aerial area.  
		
		\begin{figure}[t]
			\centering
			\includegraphics[width=4.2in]{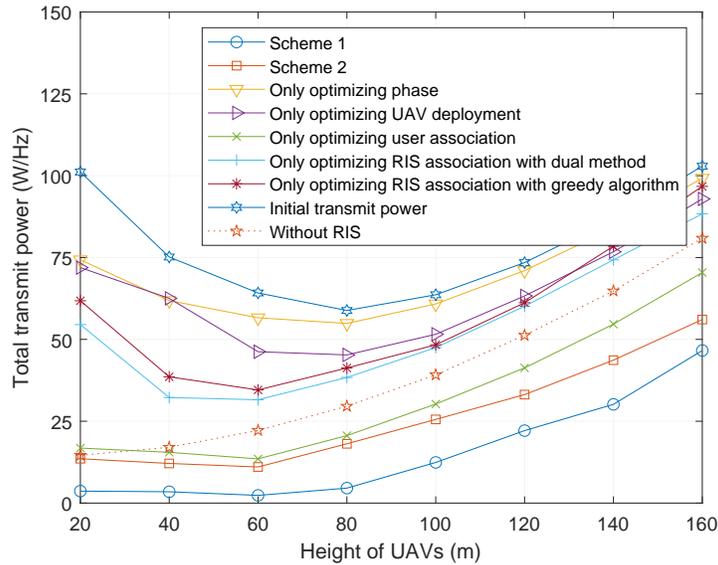}
			\caption{Total transmit power versus height of UAVs $H$.} \label{height}
		\end{figure}
	
	Fig. \ref{height} shows that  the total transmit power of all the UAVs versus the height of UAVs $H$. It is interesting to find that the power required by the two proposed schemes  decreases before it increases. And when the height is around $60 \text{m} $, the total transmit power achieves the minimum value. In order to find out the reasons behind this phenomenon, we also plot a line that reflects the transmit power changes as the UAVs fly higher when no RISs are deployed in the same region. We can see that this line increases monotonely, which indicates that the existence of the UAV-RIS-user link makes the total channel gain greater first and turn to be smaller later as the height of UAV increases. This is due to in large part to the fact that when the UAVs fly from $20\text{m}$ to $60\text{m}$, the  value of cosine of emission angles gets larger. And the impact of the  increase of emission angles on the total transmit power is greater than that of the increase  of UAVs' height. However, when  the height of UAVs grows more than $60\text{m}$, the impact of the  increase of the height is greater than that of the increase  of emission angles. When there exists no RISs, only the height of UAVs has an influence on the total transmit power of all the UAVs. Hence, in this case, the total transmit power will always increase as the height increases. Futhermore, the four schemes that optimizes part of the variables all emerge similar trends as Scheme $\text{\Rmnum{1}}$ and Scheme $\text{\Rmnum{2}}$, which can also be attributed to the  UAV-RIS-user links.   Moreover, the scheme that only optimizes user association is least affected as the height rises from $20\text{m}$ to $60\text{m}$. This is due to that by allocating ground users to a closer UAV, the emission angles of UAVs gets smaller, thus the impact of the changes of angles gets reduced. As a comparison, the total power of the scheme that only optimizes phases, in  which the locations of UAVs, RISs and users keep fixed, does not increases until $80\text{m}$.

\begin{figure}[t]
		\centering
		\includegraphics[width=4.2in]{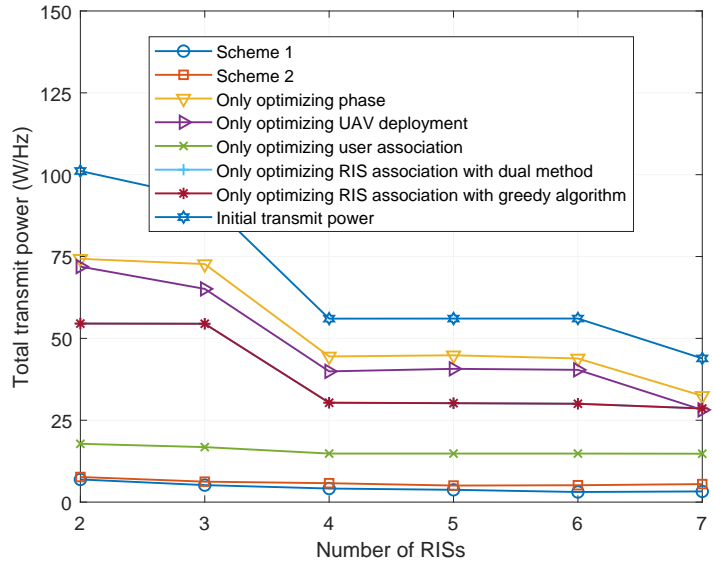}
		\caption{Total transmit power versus number of RISs $L$.} \label{L}
	\end{figure}

	Fig. \ref{L} depicts how the total transmit power of all the UAVs to meet the data rate requirement and illumination requirement of all the ground users changes as the number of RISs $L$ varies. As can be seen in Fig. \ref{L},  as the number of RISs increases in this specific area, the total transmit power of all the schemes  decreases. Since a RIS, of which the phase of each reflective elements is optimized, can constructively improve the channel state between a UAV and its associated ground users, the total transmit power will decrease correspondingly with the number of RISs increasing. Adding one RIS with $M=5$ reflective elements can yield up to $23.01\%$ and $18.58\%$ reductions in terms of the total transmit power on average through Scheme $\text{\Rmnum{1}}$ and Scheme $\text{\Rmnum{2}}$, respectively. In addition, even though we do not optimize any variables, the total transmit power  can still reduce by $21.73\%$ on average with initial all the phases equal to zero each time a RIS is added randomly.
		
	\begin{figure}[t]
		\centering
		\includegraphics[width=4.2in]{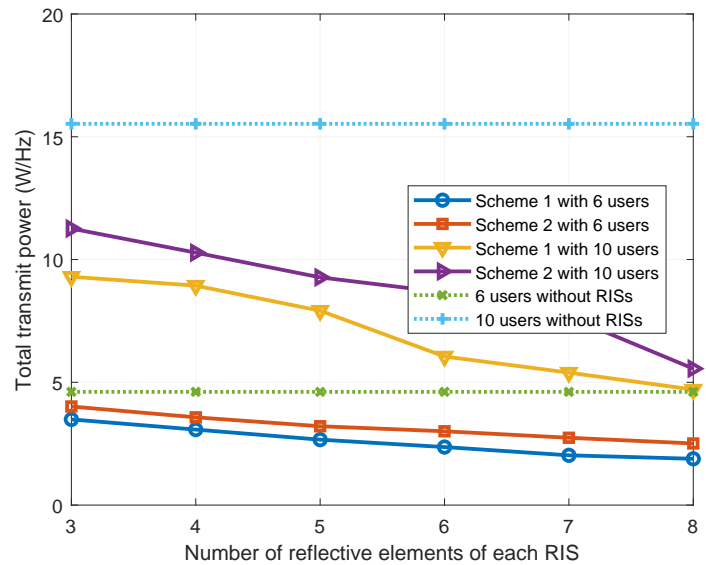}
		\caption{Total transmit power versus number of reflective elements of each RIS $M$.} \label{M}
	\end{figure}
	Fig. \ref{M} shows the total transmit power versus the number of reflective elements of each RIS $M$ with $6$ users and $10$ users, respectively. As can be seen, the total transmit power of all the UAVs decreases with the increase of the number of reflective elements, which is similar to what is shown in Fig. \ref{L}. And the proposed Scheme $\text{\Rmnum{1}}$ outperforms Scheme $\text{\Rmnum{2}}$ in terms to the total power. Meanwhile, both Scheme $\text{\Rmnum{1}}$ and  Scheme $\text{\Rmnum{2}}$ outperforms the scheme without RISs by up to $29.56\%$ and $22.79\%$, respectively, when the number of ground users is $6$.  When there are $10$ ground users,  the performance of Scheme $\text{\Rmnum{1}}$ and  Scheme $\text{\Rmnum{2}}$ yields up to $34.85\%$ and $32.11\%$ improvement, respectively. From Fig. \ref{L} and Fig. \ref{M}, we can find that the increase the number of RISs and reflective elements of each RISs both can constructively reduce the total transmit power.

	To further evaluate the computational efficiency of the two proposed schemes, Table \ref{table2} displays the running time of the two schemes in each iteration when the number of ground users is $9$ and $15$, respectively. It can be seen that   Scheme $\text{\Rmnum{1}}$ needs more time than  Scheme $\text{\Rmnum{2}}$. This is caused by the fact that Scheme $\text{\Rmnum{1}}$ searches for a global minimum point through multiple iterations, while Scheme $\text{\Rmnum{2}}$ only optimize one RISs at one time until all the RISs' associations are optimized. However, from Fig. \ref{user}, Fig. \ref{height}, Fig. \ref{L} and Fig. \ref{M}, we can find that Scheme $\text{\Rmnum{1}}$ outperforms Scheme $\text{\Rmnum{2}}$ in terms to total transmit power performance. In summary, the deployment optimized by Scheme $\text{\Rmnum{1}}$ has better power performance, which are suitable for effective communication. While Scheme $\text{\Rmnum{2}}$ is preferable for urgent communication due to its less running time.
	
	\begin{table}[htbp]
		\caption{Running Time of Two Schemes} 
		\label{table2} 
		\centering 
		\begin{tabular}{|m{1.15cm}<{\centering}|m{3.2cm}<{\centering}|m{3.2cm}<{\centering}|}
			\hline
			Number of users & Running time of Scheme $\text{\Rmnum{1}}$ in each iteration & Running time of Scheme $\text{\Rmnum{2}}$ in each iteration \\
			\hline
			$9$ & $4.649679s$ & $3.995187s$\\
			\hline
			$15$ & $11.230807s$ & $5.028470s$\\
			\hline
		\end{tabular}
	\end{table}
	
	\begin{figure}[t]
		\centering
		\includegraphics[width=4.2in]{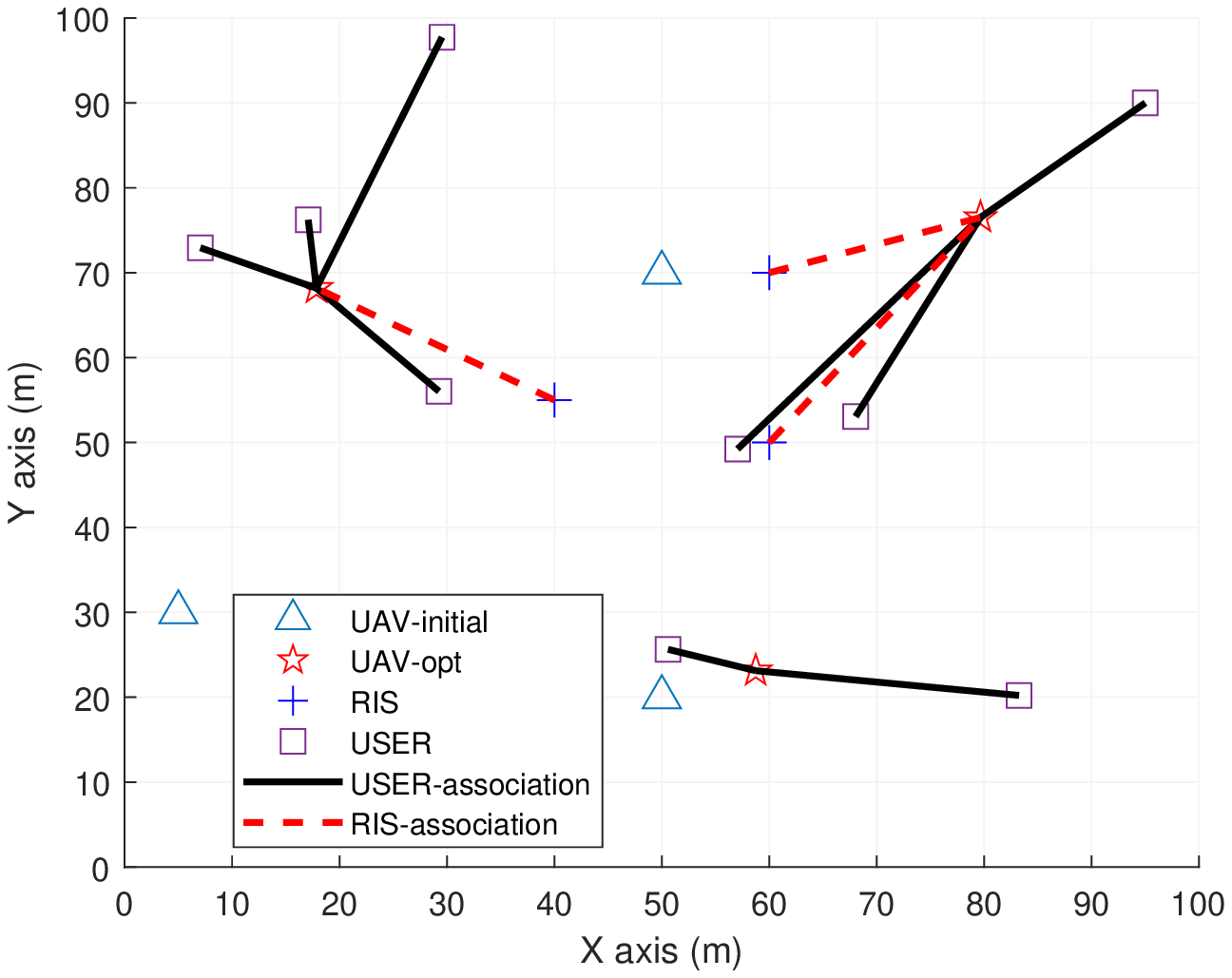}
		\caption{One example of Scheme $\text{\Rmnum{1}}$.} \label{asso1}
	\end{figure}

	\begin{figure}[t]
		\centering
		\includegraphics[width=4.2in]{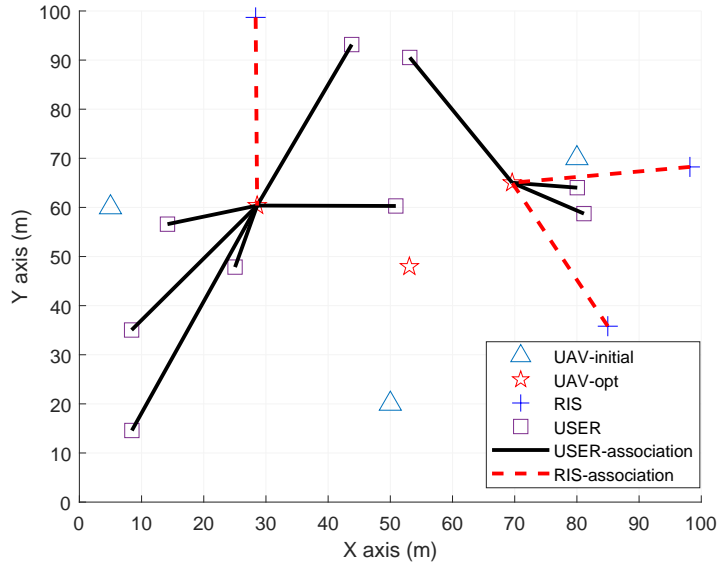}
		\caption{The other example of Scheme $\text{\Rmnum{1}}$.} \label{asso2}
	\end{figure}
	Fig. \ref{asso1} and Fig. \ref{asso2} show two illustrative examples of the proposed Scheme $\text{\Rmnum{1}}$  where the initial parameters are set randomly. In these two figures, the initial and the optimal locations of UAVs are denoted by triangles and pentacles, respectively. As can be seen, the optimal locations of UAVs are surrounded by its associated users and RISs. Futhermore, a small number of users who are distant from most users are served by an individual UAV, which can reduce the total transmit power significantly. In Fig. \ref{asso2}, however, one UAV is idle. This is because, during the period of optimization, the users who are originally associated with this UAV lead to  less total power if they are served by the other UAVs.  What can be predicted is that when ground users are close enough to each other, maybe one UAV is enough. 
	Moreover, as can be seen from Fig. \ref{asso1} and Fig. \ref{asso2}, each RIS is associated with the closest UAV among all the UAVs. This can be explained by that short distance between UAV and RIS can enhance the channel gains between UAV and ground users, thus reducing the transmit power.

	\section{conclusion}
	In this paper, we proposed a novel framework of a VLC-enabled UAV multicell network, by leveraging the prevalent RISs for reducing total consuming power of all the UAVs. 
	An optimization problem was formulated to dynamically optimize UAV deployment, phase shift of RISs and the association of ground users and RISs to achieve the  minimum total transmitting power by taking into account traffic and illumination demands. 
	An alternating algorithm was proposed to solve multiple subproblems iteratively. Numerical results demonstrate the superior performance of our proposed algorithms over the case without RIS in terms of energy consumption reduction. 
	And by associating each RIS with the closest UAV, we can achieve the minimum transmit power of all the UAVs.
	

	\bibliography{IEEEabrv,MMM}

\begin{thebibliography}{10}
\providecommand{\url}[1]{#1}
\csname url@samestyle\endcsname
\providecommand{\newblock}{\relax}
\providecommand{\bibinfo}[2]{#2}
\providecommand{\BIBentrySTDinterwordspacing}{\spaceskip=0pt\relax}
\providecommand{\BIBentryALTinterwordstretchfactor}{4}
\providecommand{\BIBentryALTinterwordspacing}{\spaceskip=\fontdimen2\font plus
\BIBentryALTinterwordstretchfactor\fontdimen3\font minus
  \fontdimen4\font\relax}
\providecommand{\BIBforeignlanguage}[2]{{%
\expandafter\ifx\csname l@#1\endcsname\relax
\typeout{** WARNING: IEEEtran.bst: No hyphenation pattern has been}%
\typeout{** loaded for the language `#1'. Using the pattern for}%
\typeout{** the default language instead.}%
\else
\language=\csname l@#1\endcsname
\fi
#2}}
\providecommand{\BIBdecl}{\relax}
\BIBdecl

\bibitem{8245811}
C.~{Kai}, H.~{Li}, L.~{Xu}, Y.~{Li}, and T.~{Jiang}, ``Energy-efficient
  device-to-device communications for green smart cities,'' \emph{IEEE
  Transactions on Industrial Informatics}, vol.~14, no.~4, pp. 1542--1551, Jan.
  2018.

\bibitem{8014294}
X.~{Chen}, L.~{Pu}, L.~{Gao}, W.~{Wu}, and D.~{Wu}, ``Exploiting massive d2d
  collaboration for energy-efficient mobile edge computing,'' \emph{IEEE
  Wireless Communications}, vol.~24, no.~4, pp. 64--71, Aug. 2017.

\bibitem{9170653}
F.~{Tariq}, M.~R.~A. {Khandaker}, K.~K. {Wong}, M.~A. {Imran}, M.~{Bennis}, and
  M.~{Debbah}, ``A speculative study on 6g,'' \emph{IEEE Wireless
  Communications}, vol.~27, no.~4, pp. 118--125, Aug. 2020.

\bibitem{8910627}
Q.~{Wu} and R.~{Zhang}, ``Towards smart and reconfigurable environment:
  Intelligent reflecting surface aided wireless network,'' \emph{IEEE
  Communications Magazine}, vol.~58, no.~1, pp. 106--112, Nov. 2020.

\bibitem{pan2020reconfigurable}
C.~Pan, H.~Ren, K.~Wang, M.~Elkashlan, M.~Chen, M.~Di~Renzo, Y.~Hao, J.~Wang,
  A.~L. Swindlehurst, X.~You \emph{et~al.}, ``Reconfigurable intelligent
  surface for 6g and beyond: Motivations, principles, applications, and
  research directions,'' \emph{arXiv preprint arXiv:2011.04300}, Nov. 2020.

\bibitem{9090356}
C.~{Pan}, H.~{Ren}, K.~{Wang}, W.~{Xu}, M.~{Elkashlan}, A.~{Nallanathan}, and
  L.~{Hanzo}, ``Multicell mimo communications relying on intelligent reflecting
  surfaces,'' \emph{IEEE Transactions on Wireless Communications}, vol.~19,
  no.~8, pp. 5218--5233, May 2020.

\bibitem{9201413}
S.~{Zeng}, H.~{Zhang}, B.~{Di}, Z.~{Han}, and L.~{Song}, ``Reconfigurable
  intelligent surface (ris) assisted wireless coverage extension: Ris
  orientation and location optimization,'' \emph{IEEE Communications Letters},
  pp. 1--1, Sep. 2020.

\bibitem{yang2020beamforming}
Z.~Yang, W.~Xu, C.~Huang, J.~Shi, and M.~Shikh-Bahaei, ``Beamforming design for
  multiuser transmission through reconfigurable intelligent surface,''
  \emph{IEEE Transactions on Communications}, Oco. 2020.

\bibitem{9027303}
V.~C. {Thirumavalavan} and T.~S. {Jayaraman}, ``Ber analysis of reconfigurable
  intelligent surface assisted downlink power domain noma system,'' in
  \emph{Proc. of International Conference on COMmunication Systems NETworkS
  (COMSNETS)}, March 2020, pp. 519--522.

\bibitem{9247583}
L.~{Yang} and Y.~{Yuan}, ``Secrecy outage probability analysis for ris-assisted
  noma systems,'' \emph{Electronics Letters}, vol.~56, no.~23, pp. 1254--1256,
  Nov. 2020.

\bibitem{9217565}
M.~{Elhattab}, M.~A. {Arfaoui}, C.~{Assi}, and A.~{Ghrayeb}, ``Reconfigurable
  intelligent surface assisted coordinated multipoint in downlink noma
  networks,'' \emph{IEEE Communications Letters}, pp. 1--1, Oco. 2020.

\bibitem{9217212}
M.~{Zhang}, M.~{Chen}, Z.~{Yang}, H.~{Asgari}, and M.~{Shikh-Bahaei}, ``Joint
  user clustering and passive beamforming for downlink noma system with
  reconfigurable intelligent surface,'' in \emph{Proc. of IEEE Annual
  International Symposium on Personal, Indoor and Mobile Radio Communications},
  London, United Kingdom, United Kingdom, Oco. 2020, pp. 1--6.

\bibitem{9133094}
T.~{Hou}, Y.~{Liu}, Z.~{Song}, X.~{Sun}, Y.~{Chen}, and L.~{Hanzo},
  ``Reconfigurable intelligent surface aided noma networks,'' \emph{IEEE
  Journal on Selected Areas in Communications}, vol.~38, no.~11, pp.
  2575--2588, July 2020.

\bibitem{9217117}
Y.~{Xu}, M.~{Chen}, Z.~{Yang}, Y.~{Liu}, H.~{Long}, and M.~{Shikh-Bahaei},
  ``Fair non-orthogonal multiple access communication systems with
  reconfigurable intelligent surface,'' in \emph{Proc. of IEEE Annual
  International Symposium on Personal, Indoor and Mobile Radio Communications},
  London, United Kingdom, United Kingdom, Oco. 2020, pp. 1--6.

\bibitem{9234527}
Y.~{Li}, M.~{Jiang}, Q.~{Zhang}, and J.~{Qin}, ``Joint beamforming design in
  multi-cluster miso noma reconfigurable intelligent surface-aided downlink
  communication networks,'' \emph{IEEE Transactions on Communications}, pp.
  1--1, Oco. 2020.

\bibitem{9148537}
M.~{Jian} and Y.~{Zhao}, ``A modified off-grid sbl channel estimation and
  transmission strategy for ris-assisted wireless communication systems,'' in
  \emph{Proc. of International Wireless Communications and Mobile Computing
  (IWCMC)}, July 2020, pp. 1848--1853.

\bibitem{9224676}
M.~{Nemati}, J.~{Park}, and J.~{Choi}, ``Ris-assisted coverage enhancement in
  millimeter-wave cellular networks,'' \emph{IEEE Access}, vol.~8, pp.
  188\,171--188\,185, Oco. 2020.

\bibitem{9110888}
X.~{Yang}, C.~K. {Wen}, and S.~{Jin}, ``Mimo detection for reconfigurable
  intelligent surface-assisted millimeter wave systems,'' \emph{IEEE Journal on
  Selected Areas in Communications}, vol.~38, no.~8, pp. 1777--1792, June 2020.

\bibitem{9148781}
N.~S. {Perović}, M.~D. {Renzo}, and M.~F. {Flanagan}, ``Channel capacity
  optimization using reconfigurable intelligent surfaces in indoor mmwave
  environments,'' in \emph{Proc. of ICC IEEE International Conference on
  Communications (ICC)}, July 2020, pp. 1--7.

\bibitem{9238887}
J.~{Zhang}, Z.~{Zheng}, Z.~{Fei}, and X.~{Bao}, ``Positioning with dual
  reconfigurable intelligent surfaces in millimeter-wave mimo systems,'' in
  \emph{Proc. of IEEE/CIC International Conference on Communications in China
  (ICCC)}, Nov. 2020, pp. 800--805.

\bibitem{6685758}
L.~{Grobe}, A.~{Paraskevopoulos}, J.~{Hilt}, D.~{Schulz}, F.~{Lassak},
  F.~{Hartlieb}, C.~{Kottke}, V.~{Jungnickel}, and K.~{Langer}, ``High-speed
  visible light communication systems,'' \emph{IEEE Communications Magazine},
  vol.~51, no.~12, pp. 60--66, Dec. 2013.

\bibitem{6963803}
S.~{Wu}, H.~{Wang}, and C.~{Youn}, ``Visible light communications for 5g
  wireless networking systems: from fixed to mobile communications,''
  \emph{IEEE Network}, vol.~28, no.~6, pp. 41--45, Nov. 2014.

\bibitem{6072221}
J.~J.~D. {McKendry}, D.~{Massoubre}, S.~{Zhang}, B.~R. {Rae}, R.~P. {Green},
  E.~{Gu}, R.~K. {Henderson}, A.~E. {Kelly}, and M.~D. {Dawson},
  ``Visible-light communications using a cmos-controlled micro-light-
  emitting-diode array,'' \emph{Journal of Lightwave Technology}, vol.~30,
  no.~1, pp. 61--67, Nov. 2012.

\bibitem{7239528}
P.~H. {Pathak}, X.~{Feng}, P.~{Hu}, and P.~{Mohapatra}, ``Visible light
  communication, networking, and sensing: A survey, potential and challenges,''
  \emph{IEEE Communications Surveys Tutorials}, vol.~17, no.~4, pp. 2047--2077,
  Sep. 2015.

\bibitem{7072557}
D.~{Karunatilaka}, F.~{Zafar}, V.~{Kalavally}, and R.~{Parthiban}, ``Led based
  indoor visible light communications: State of the art,'' \emph{IEEE
  Communications Surveys Tutorials}, vol.~17, no.~3, pp. 1649--1678, Mar. 2015.

\bibitem{8715400}
Y.~{Yang}, M.~{Chen}, C.~{Guo}, C.~{Feng}, and W.~{Saad}, ``Power efficient
  visible light communication with unmanned aerial vehicles,'' \emph{IEEE
  Communications Letters}, vol.~23, no.~7, pp. 1272--1275, May 2019.

\bibitem{9140367}
Y.~{Wang}, M.~{Chen}, Z.~{Yang}, T.~{Luo}, and W.~{Saad}, ``Deep learning for
  optimal deployment of uavs with visible light communications,'' \emph{IEEE
  Transactions on Wireless Communications}, vol.~19, no.~11, pp. 7049--7063,
  July 2020.

\bibitem{9075277}
Q.~V. {Pham}, T.~{Huynh-The}, M.~{Alazab}, J.~{Zhao}, and W.~J. {Hwang},
  ``Sum-rate maximization for uav-assisted visible light communications using
  noma: Swarm intelligence meets machine learning,'' \emph{IEEE Internet of
  Things Journal}, vol.~7, no.~10, pp. 10\,375--10\,387, Apr. 2020.

\bibitem{5682214}
K.~{Lee}, H.~{Park}, and J.~R. {Barry}, ``Indoor channel characteristics for
  visible light communications,'' \emph{IEEE Communications Letters}, vol.~15,
  no.~2, pp. 217--219, Jan. 2011.

\bibitem{8301878}
C.~{Huang} and X.~{Zhang}, ``Los-nlos identification algorithm for indoor
  visible light positioning system,'' in \emph{Proc. of International Symposium
  on Wireless Personal Multimedia Communications (WPMC)}, Bali, Indonesia, Dec.
  2017, pp. 575--578.

\end{thebibliography}

\end{document}